    \documentclass[12pt,a4paper]{article} 
    \usepackage{psfig}  \usepackage{epsfig}  \usepackage{graphics}
\def\beq{\begin{equation}} 
\def\eeq{\end{equation}} 
\def\bea{\begin{eqnarray}} 
\def\eea{\end{eqnarray}}
\def\bq{\begin{quote}} 
\def\eq{\end{quote}}

\def\gappeq{\mathrel{\rlap {\raise.5ex\hbox{$>$}} {\lower.5ex\hbox{$\sim$}}}}

\def\lappeq{\mathrel{\rlap{\raise.5ex\hbox{$<$}} {\lower.5ex\hbox{$\sim$}}}}

\def\lsim{\mathrel{\rlap{\lower4pt\hbox{\hskip1pt$\sim$}}
    \raise1pt\hbox{$<$}}}                
\def\gsim{\mathrel{\rlap{\lower4pt\hbox{\hskip1pt$\sim$}}
    \raise1pt\hbox{$>$}}}                
\begin{document}
\vspace*{10mm}
\begin{center}  \begin{Large} \begin{bf}
The Standard Electroweak Theory and Beyond\\
  \end{bf}  \end{Large}
  \bigskip
  \bigskip
  \begin{large}
\renewcommand{\thefootnote}{\fnsymbol{footnote}}
\setcounter{footnote}{0}
G. Altarelli \footnote[1]{{\it Electronic mail address:} 
{\tt Guido.Altarelli@cern.ch}}\\
\setcounter{footnote}{0}
  \end{large}
\medskip
Theoretical Physics Division, CERN\\
1211 Geneva 23,
Switzerland\\ 
\end{center}
\bigskip
\begin{center}
{\bf Abstract}\end{center}
\begin{quotation}
\noindent
1. Introduction\\ 
2. Gauge Theories\\ 
3. The Standard Model of Electroweak
Interactions\\
4. The Higgs Mechanism\\
5. The CKM Matrix\\
6. Renormalisation and Higher Order Corrections\\
7. Why we do
Believe in the Standard Model: Precision Tests\\
\indent 7.1.Precision Electroweak Data and the Standard Model\\
\indent 7.2.A More General Analysis of Electroweak Data \\
8. Why we do not Believe in the Standard Model\\
\indent 8.1.Conceptual Problems\\
\indent 8.2.Hints from Experiment\\
\indent -8.2.1 Unification of Couplings\\
\indent -8.2.2 Dark Matter\\
\indent -8.2.3 Neutrino Masses\\
\indent -8.2.4 Baryogenesis\\
9. Status of the Search for the Higgs and for New Physics\\
10.Conclusion
\end{quotation}


\section{Introduction}

These lectures on electroweak (EW) interactions start with a short summary of the Glashow--Weinberg--Salam theory 
and then cover in detail some main subjects of present interest in phenomenology.

The modern EW theory inherits the phenomenological successes of the $(V-A) \otimes (V-A)$ four-fermion
low-energy description of weak interactions, and provides a well-defined and consistent theoretical framework
including weak interactions and quantum electrodynamics in a unified picture.

 As an introduction, we recall some salient physical features of the weak interactions. The weak
interactions derive their name from their intensity. At low energy the strength of the effective four-fermion
interaction of charged currents is determined by the Fermi coupling constant $G_F$. For example, the effective
interaction for muon decay is given by
\beq {\cal L}_{\rm eff} = (G_F/\sqrt 2) \left[ \bar
\nu_{\mu}\gamma_{\alpha}(1-\gamma_5)\mu \right]
\left[ \bar e\gamma^{\alpha}(1-\gamma_5)\nu_e \right]~,
\label{1}
\eeq with \cite{pdg} \footnote{For reasons of space, here only a few basic references are listed. Starting from those a more
extended bibliography can easily be found.}
\beq G_F = 1.16639(1) \times 10^{-5}~{\rm GeV}^{-2}~.
\label{2}
\eeq In natural units $ \hbar = c = 1$, $G_F$ has dimensions of (mass)$^{-2}$. As a result, the intensity of weak
interactions at low energy is characterized by
$G_FE^2$, where $E$ is the energy scale for a given process ($E \approx m_{\mu}$  for muon decay). Since
\begin{equation} G_FE^2 = G_Fm^2_p(E/m_p)^2 \simeq 10^{-5}(E/m_p)^2~,
\label{3}
\end{equation} where $m_p$ is the proton mass, the weak interactions are indeed weak at low energies (energies of order
$m_p$). Effective four
fermion couplings for neutral current interactions have comparable intensity and energy behaviour. The quadratic increase with
energy cannot continue for ever, because it would lead to a violation of unitarity. In fact, at large energies the propagator
effects can no longer be neglected, and the current--current interaction is resolved into current--$W$ gauge boson vertices
connected by a $W$ propagator. The strength of the weak interactions at high energies is then measured by $g_W$, the
$W-\mu$--$\nu_{\mu}$ coupling, or, even better, by
$\alpha_W = g^2_W/4\pi$ analogous to the fine-structure constant $\alpha$ of QED. In the standard EW theory, we have
\begin{equation}
\alpha_W = \sqrt 2~G_F~m^2_W/\pi = \alpha/\sin^2\theta_W \cong 1/30~.
\label{4}
\end{equation} That is, at high energies the weak interactions are no longer so weak.

 The range $r_W$ of weak interactions is very short: it is only with the experimental discovery of the $W$ and $Z$
gauge bosons that it could be demonstrated that $r_W$ is non-vanishing. Now we know that
\begin{equation} r_W = \hbar /m_Wc \simeq 2.5 \times 10^{-16}~{\rm cm}~,
\label{5}
\end{equation} corresponding to $m_W \simeq 80$~GeV. This very large value for the $W$ (or the
$Z$) mass makes a drastic difference, compared with the massless photon and the infinite range of the QED force. The
direct experimental limit on the photon mass is \cite{pdg}
$m_{\gamma} <2~10^{-16}~eV$. Thus, on the one hand, there is very good evidence that the photon is massless. On the
other hand, the weak bosons are very heavy. A unified theory of EW interactions has to face this striking
difference.

Another apparent obstacle in the way of EW unification is the chiral structure of weak interactions: in the
massless limit for fermions, only left-handed quarks and leptons (and right-handed antiquarks and antileptons) are
coupled to $W$'s. This clearly implies parity and charge-conjugation violation in weak interactions.

The universality of weak interactions and the algebraic properties of the electromagnetic and weak currents [the
conservation of vector currents (CVC), the partial conservation of axial currents (PCAC), the algebra of currents,
etc.] have been crucial in pointing to a symmetric role of electromagnetism and weak interactions at a more fundamental
level. The old Cabibbo universality for the weak charged current:
\begin{eqnarray} J^{\rm weak}_{\alpha} &=&
\bar \nu_{\mu}\gamma_{\alpha} (1-\gamma_5)\mu +
\bar \nu_e\gamma_{\alpha}(1-\gamma_5) e +
\cos\theta_c~\bar u \gamma_{\alpha}(1-\gamma_5)d + \nonumber \\  &&\sin \theta_c~\bar u \gamma_{\alpha}(1-\gamma_5)s +
...~,
\label{6}
\end{eqnarray} suitably extended, is naturally implied by the standard EW theory. In this theory the weak
gauge bosons couple to all particles with couplings that are proportional to their weak charges, in the same way as the
photon couples to all particles in proportion to their electric charges [in Eq.~(\ref{6}), $d' =
\cos\theta_c~d + \sin \theta_c~s$ is the weak-isospin partner of $u$ in a doublet.  The $(u,d')$ doublet has the same
couplings as the $(\nu_e,\ell)$ and 
$(\nu_{\mu},\mu)$ doublets].

 Another crucial feature is that the charged weak interactions are the only known interactions that can change flavour:
charged leptons into neutrinos or up-type quarks into down-type quarks. On the contrary, there are no flavour-changing
neutral currents at tree level. This is a remarkable property of the weak neutral current, which is explained by the
introduction of the Glashow-Iliopoulos-Maiani mechanism and has led to the successful prediction of charm.

 The natural suppression of flavour-changing neutral currents, the separate conservation of $e, \mu$  and $\tau$
leptonic flavours, the mechanism of CP violation through the phase in the quark-mixing matrix, are all crucial
features of the Standard Model. Many examples of new physics tend to break the selection rules of the standard theory.
Thus the experimental study of rare flavour-changing transitions is an important window on possible new physics.

 In the following sections we shall see how these properties of weak interactions fit into the standard EW
theory.

\section{Gauge Theories}

In this section we summarize the definition and the structure of a gauge Yang--Mills theory. We
will list here the general rules for constructing such a theory. Then in the next section these results will be applied
to the EW theory.

Consider a Lagrangian density ${\cal L}[\phi,\partial_{\mu}\phi]$ which is invariant under a $D$ dimensional continuous
group of transformations:
\begin{equation}
\phi' = U(\theta^A)\phi\quad\quad (A = 1, 2, ..., D)~.
\label{7}
\end{equation} For $\theta^A$ infinitesimal, $U(\theta^A) = 1 + ig \sum_A~\theta^AT^A$, where
$T^A$ are the generators of the group $\Gamma$ of transformations (\ref{7}) in the (in general reducible)
representation of the fields $\phi$. Here we restrict ourselves to the case of internal symmetries, so that $T^A$ are
matrices that are independent of the space--time coordinates. The generators $T^A$ are normalized in such a way that
for the lowest dimensional non-trivial representation of the group $\Gamma$ (we use $t^A$ to denote the generators in
this particular representation) we have
\begin{equation} {\rm tr}(t^At^B) = \frac{1}{2} \delta^{AB}~.
\label{8}
\end{equation} The generators satisfy the commutation relations
\begin{equation} [T^A,T^B] = iC_{ABC}T^C~.
\label{9}
\end{equation} In the following, for each quantity $V^A$ we define
\begin{equation} {\bf V} = \sum_A~T^AV^A~.
\label{10}
\end{equation} If we now make the parameters $\theta^A$ depend on the space--time coordinates
$\theta^A = \theta^A(x_{\mu}),$ ${\cal L}[\phi,\partial_{\mu}\phi]$ is in general no longer invariant under the gauge
transformations $U[\theta^A(x_{\mu})]$, because of the derivative terms. Gauge invariance is recovered if the ordinary
derivative is replaced by the covariant derivative:
\begin{equation} D_{\mu} = \partial_{\mu} + ig{\bf V}_{\mu}~,
\label{11}
\end{equation} where $V^A_{\mu}$ are a set of $D$ gauge fields (in one-to-one correspondence with the group generators)
with the transformation law
\begin{equation} {\bf V}'_{\mu} = U{\bf V}_{\mu}U^{-1} - (1/ig)(\partial_{\mu}U)U^{-1}~.
\label{12}
\end{equation} For constant $\theta^A$, {\bf V} reduces to a tensor of the adjoint (or regular) representation of the
group:
\begin{equation} {\bf V}'_{\mu} = U{\bf V}_{\mu}U^{-1} \simeq {\bf V}_{\mu} + ig[\theta, {\bf V}_{\mu}]~,
\label{13}
\end{equation} which implies that
\begin{equation} V'^C_{\mu} = V^C_{\mu} - gC_{ABC}\theta^AV^B_{\mu}~,
\label{14}
\end{equation} where repeated indices are summed up.

As a consequence of Eqs. (\ref{11}) and (\ref{12}), $D_{\mu}\phi$  has the same transformation pro\-perties as $\phi$:
\begin{equation} (D_{\mu}\phi)' = U(D_{\mu}\phi)~.
\label{15}
\end{equation}

Thus ${\cal L}[\phi,D_{\mu}\phi]$ is indeed invariant under gauge transformations. In order to construct a
gauge-invariant kinetic energy term for the gauge fields $V^A$, we consider
\begin{equation} [D_{\mu},D_{\nu}] \phi =  ig\{\partial_{\mu}{\bf V}_{\nu} - \partial_{\nu}{\bf V}_{\mu} + ig[{\bf
V}_{\mu},{\bf V}_{\nu}]\}\phi \equiv ig {\bf F}_{\mu\nu}\phi~,
\label{16}
\end{equation} which is equivalent to
\begin{equation} F^A_{\mu\nu} = \partial_{\mu}V^A_{\nu} - \partial_{\nu}V^A_{\mu} - gC_{ABC}V^B_{\mu}V^C_{\nu}~.
\label{17}
\end{equation} From Eqs. (\ref{7}), (\ref{15}) and (\ref{16}) it follows that the transformation properties of
$F^A_{\mu\nu}$ are those of a tensor of the adjoint representation
\begin{equation} {\bf F}'_{\mu\nu} = U{\bf F}_{\mu\nu}U^{-1}~.
\label{18}
\end{equation} The complete Yang--Mills Lagrangian, which is invariant under gauge transformations, can be written in
the form
\begin{equation} {\cal L}_{\rm YM} = - \frac{1}{4} \sum_A F^A_{\mu\nu}F^{A\mu\nu} + {\cal L} [\phi,D_{\mu}\phi]~.
\label{19}
\end{equation}

For an Abelian theory, as for example QED, the gauge transformation reduces to
$U[\theta(x)] = {\rm exp} [ieQ\theta(x)]$, where $Q$ is the charge generator. The associated gauge field (the photon),
according to Eq. (\ref{12}), transforms as
\begin{equation} V'_{\mu} = V_{\mu} - \partial_{\mu}\theta(x)~.
\label{20}
\end{equation} In this case, the $F_{\mu\nu}$ tensor is linear in the gauge field $V_{\mu}$ so that in the absence of
matter fields the theory is free. On the other hand, in the non-Abelian case the $F^A_{\mu\nu}$ tensor contains both
linear and quadratic terms in $V^A_{\mu}$, so that the theory is non-trivial even in the absence of matter fields.

\section{The Standard Model of Electroweak Interactions}

 In this section, we summarize the structure of the standard EW Lagrangian and specify the couplings of
$W^{\pm}$ and $Z$, the intermediate vector bosons. 

For this discussion we split the Lagrangian into two parts by separating the Higgs boson couplings:
\begin{equation} {\cal L} = {\cal L}_{\rm symm} + {\cal L}_{\rm Higgs}~.
\label{21}
\end{equation}

We start by specifying ${\cal L}_{\rm symm}$, which involves only gauge bosons and fermions:
\begin{eqnarray} {\cal L}_{\rm symm} &=& -\frac{1}{4}~\sum^3_{A=1}~F^A_{\mu\nu}F^{A\mu\nu} -
\frac{1}{4}B_{\mu\nu}B^{\mu\nu} +
\bar\psi_Li\gamma^{\mu}D_{\mu}\psi_L \nonumber \\ &&+  \bar\psi_Ri\gamma^{\mu}D_{\mu}\psi_R~.
\label{22}
\end{eqnarray} This is the Yang--Mills Lagrangian for the gauge group $SU(2)\otimes U(1)$ with fermion matter fields.
Here
\begin{equation} B_{\mu\nu}  =  \partial_{\mu}B_{\nu} - \partial_{\nu}B_{\mu} \quad {\rm and} \quad F^A_{\mu\nu} =
\partial_{\mu}W^A_{\nu} - \partial_{\nu}W^A_{\mu}  - g \epsilon_{ABC}~W^B_{\mu}W^C_{\nu}
\label{23}
\end{equation} are the gauge antisymmetric tensors constructed out of the gauge field $B_{\mu}$ associated with $U(1)$,
and $W^A_{\mu}$ corresponding to the three $SU(2)$ generators; $\epsilon_{ABC}$ are the group structure constants [see
Eqs. (\ref{9})] which, for $SU(2)$, coincide with the totally antisymmetric Levi-Civita tensor (recall the familiar
angular momentum commutators). The normalization of the $SU(2)$ gauge coupling $g$ is therefore specified by
Eq.~(\ref{23}).

The fermion fields are described through their left-hand and right-hand components:
\begin{equation}
\psi_{L,R} = [(1 \mp \gamma_5)/2]\psi, \quad
\bar \psi_{L,R} = \bar \psi[(1 \pm \gamma_5)/2]~,
\label{24}
\end{equation} with $\gamma_5$ and other Dirac matrices defined as in the book by Bjorken--Drell. In particular, $\gamma^2_5
= 1, \gamma_5^{\dag} = \gamma_5$. Note that, as given in Eq. (\ref{24}),
$$
\bar\psi_L = 
\psi^{\dag}_L\gamma_0 = \psi^{\dag}[(1-\gamma_5)/2]\gamma_0 =
\bar\psi[\gamma_0(1-\gamma_5)/2]\gamma_0 = \bar \psi[(1 + \gamma_5)/2]~.
$$ The matrices $P_{\pm} = (1 \pm \gamma_5)/2$ are projectors. They satisfy the relations $P_{\pm}P_{\pm} = P_{\pm},
P_{\pm}P_{\mp} = 0, P_+ + P_- = 1$.

The sixteen linearly independent Dirac matrices can be divided into
$\gamma_5$-even and $\gamma_5$-odd according to whether they commute or anticommute with $\gamma_5$. For the
$\gamma_5$-even, we have
\begin{equation}
\bar \psi\Gamma_E \psi = \bar \psi_L\Gamma_E\psi_R + \bar \psi_R\Gamma_E\psi_L
\quad\quad (\Gamma_E \equiv 1, i\gamma_5, \sigma_{\mu\nu})~,
\label{25}
\end{equation} whilst for the $\gamma_5$-odd,
\begin{equation}
\bar \psi\Gamma_O \psi = \bar \psi_L\Gamma_O\psi_L + \bar \psi_R\Gamma_O\psi_R
\quad\quad (\Gamma_O \equiv \gamma_{\mu}, \gamma_{\mu}\gamma_5)~.
\label{26}
\end{equation} In the Standard Model (SM) the left and right fermions have different transformation properties under
the gauge group. Thus, mass terms for fermions (of the form
$\bar\psi_L\psi_R$ + h.c.) are forbidden in the symmetric limit. In particular, all $\psi_R$ are singlets in the
Minimal Standard Model (MSM). But for the moment, by
$\psi_R$ we mean a column vector, including all fermions in the theory that span a generic reducible representation of
$SU(2) \otimes U(1)$. The standard EW theory is a chiral theory, in the sense that $\psi_L$ and $\psi_R$ behave
differently under the gauge group. In the absence of mass terms, there are only vector and axial vector interactions in
the Lagrangian that have the property of not mixing $\psi_L$ and $\psi_R$. Fermion masses will be introduced, together
with
$W^{\pm}$ and $Z$ masses, by the mechanism of symmetry breaking. The covariant derivatives $D_{\mu}\psi_{L,R}$ are
explicitly given by
\begin{equation} D_{\mu}\psi_{L,R} = 
\left[ \partial_{\mu} + ig \sum^3_{A=1}~t^A_{L,R}W^A_{\mu} + ig'\frac{1}{2}Y_{L,R}B_{\mu} \right] \psi_{L,R}~,
\label{27}
\end{equation}  where $t^A_{L,R}$ and $1/2Y_{L,R}$ are the $SU(2)$ and $U(1)$ generators, respectively, in the
reducible representations $\psi_{L,R}$. The commutation relations of the $SU(2)$ generators are given by
\begin{equation} [t^A_L,t^B_L] = i~\epsilon_{ABC}t^C_L \quad {\rm and} \quad [t^A_R,t^B_R] = i \epsilon_{ABC}t^C_R~.
\label{28}
\end{equation} We use the normalization (\ref{8}) [in the fundamental representation of
$SU(2)$]. The electric charge generator $Q$ (in units of $e$, the positron charge) is given by
\begin{equation} Q = t^3_L + 1/2~Y_L = t^3_R + 1/2~Y_R~.
\label{29}
\end{equation} Note that the normalization of the $U(1)$ gauge coupling $g'$ in (\ref{27}) is now specified as a
consequence of (\ref{29}).

All fermion couplings to the gauge bosons can be derived directly from Eqs. (\ref{22}) and (\ref{27}). The
charged-current (CC) couplings are the simplest. From
\begin{eqnarray} g(t^1W^1_{\mu} + t^2W^2_{\mu}) &=& g \left\{ [(t^1 + it^2)/ \sqrt 2] (W^1_{\mu} - iW^2_{\mu})/\sqrt 2]
+ {\rm h.c.} \right\}\nonumber \\
 &= &g \left\{[(t^+W^-_{\mu})/\sqrt 2] + {\rm h.c.} \right\}~,
\label{30}
\end{eqnarray} where $t^{\pm}  = t^1 \pm it^2$ and $W^{\pm} = (W^1 \pm iW^2)/\sqrt 2$, we obtain the vertex
\begin{equation} V_{\bar \psi \psi W}  =  g \bar \psi \gamma_{\mu}\left[ (t^+_L/ \sqrt 2)(1 - \gamma_5)/2 + (t^+_R/
\sqrt 2)(1 + \gamma_5)/2 \right]
 \psi W^-_{\mu} + {\rm h.c.}
\label{31}
\end{equation}

In the neutral-current (NC) sector, the photon $A_{\mu}$ and the mediator
$Z_{\mu}$ of the weak NC are orthogonal and normalized linear combinations of
$B_{\mu}$ and $W^3_{\mu}$:
\begin{eqnarray} A_{\mu} &=& \cos \theta_WB_{\mu} + \sin \theta_WW^3_{\mu}~, \nonumber \\  Z_{\mu} &=& -\sin
\theta_WB_{\mu} + \cos \theta_W~W^3_{\mu}~.
\label{32}
\end{eqnarray} Equations (\ref{32}) define the weak mixing angle $\theta_W$. The photon is characterized by equal
couplings to left and right fermions with a strength equal to the electric charge. Recalling Eq. (\ref{29}) for the
charge matrix $Q$, we immediately obtain
\begin{equation} g~\sin \theta_W = g'\cos \theta_W = e~,
\label{33}
\end{equation} or equivalently,
\begin{equation} {\rm tg}~\theta_W = g'/g
\label{34}
\end{equation} Once $\theta_W$ has been fixed by the photon couplings, it is a simple matter of algebra to derive the
$Z$ couplings, with the result
\begin{equation}
\Gamma_{\bar \psi \psi Z} = g/(2~\cos \theta_W) \bar \psi \gamma_{\mu}
  [t^3_L(1-\gamma_5) + t^3_R(1+\gamma_5) - 2Q \sin^2\theta_W] \psi Z^{\mu}~,
\label{35}
\end{equation}  where $\Gamma_{\bar \psi \psi Z}$ is a notation for the vertex. In the MSM, $t^3_R = 0$ and $t^3_L =
\pm 1/2$. 

In order to derive the effective four-fermion interactions that are equivalent, at low energies, to the CC and NC
couplings given in Eqs. (\ref{31}) and (\ref{35}), we anticipate that large masses, as experimentally observed, are
provided for $W^{\pm}$  and $Z$ by ${\cal L}_{\rm Higgs}$.  For left--left CC couplings, when the momentum transfer
squared can be neglected with respect to
$m^2_W$ in the propagator of Born diagrams with single $W$ exchange, from Eq.~(\ref{31}) we can write
 \begin{equation} {\cal L}^{\rm CC}_{\rm eff} \simeq (g^2/8m^2_W) [ \bar \psi \gamma_{\mu}(1 - \gamma_5)t^+_L\psi][
\bar \psi
\gamma^{\mu}(1 - \gamma_5) t^-_L\psi]~.
\label{36}
\end{equation}  By specializing further in the case of doublet fields such as $\nu_e-e^-$ or $
\nu_{\mu} - \mu^-$, we obtain the tree-level relation of $g$ with the Fermi coupling constant $G_F$ measured from $\mu$
decay [see Eq. (\ref{2})]:
\begin{equation}
 G_F/\sqrt 2 = g^2/8m^2_W~.
\label{37}
\end{equation} By recalling that $g~\sin \theta_W = e$, we can also cast this relation in the form
\begin{equation} m_W = \mu_{\rm Born}/ \sin \theta_W~,
\label{38}
\end{equation} with
\begin{equation}
\mu_{\rm Born} = (\pi \alpha / \sqrt 2 G_F)^{1/2} \simeq 37.2802~{\rm GeV}~,
\label{39}
\end{equation} where $\alpha$ is the fine-structure constant of QED $(\alpha \equiv e^2/4\pi = 1/137.036)$. 

In the same way, for neutral currents we obtain in Born approximation from Eq.~(\ref{35}) the effective four-fermion
interaction given by
\begin{equation} {\cal L}^{\rm NC}_{\rm eff} \simeq \sqrt 2~G_F \rho_0\bar \psi \gamma_{\mu}[...]
\psi \bar \psi \gamma^{\mu}[...] \psi~,
\label{40}
\end{equation} where
\begin{equation} [...] \equiv t^3_L(1 - \gamma_5) + t^3_R (1 + \gamma_5) - 2Q \sin^2\theta_W
\label{41}
\end{equation} and
\begin{equation}
\rho_0 = m^2_W/m^2_Z~\cos^2 \theta_W~.
\label{42}
\end{equation}

All couplings given in this section are obtained at tree level and are modified in higher orders of perturbation
theory. In particular, the relations between
$m_W$ and $\sin \theta_W$  [Eqs. (\ref{38}) and (\ref{39})] and the observed values of $\rho~(\rho = \rho_0$ at tree
level) in different NC processes, are altered by computable EW radiative corrections, as discussed in Section
6. 

The gauge-boson self-interactions can be derived from the
$F_{\mu\nu}$ term in ${\cal L}_{\rm symm}$, by using Eq. (\ref{32}) and
$W^{\pm} = (W^1 \pm iW^2)/\sqrt 2$. Defining the three-gauge-boson vertex as in Fig. 1, we obtain $(V \equiv \gamma,Z)$
\begin{equation}
\Gamma_{W^-W^+V} = ig_{W^-W^+V}[g_{\mu\nu}(q-p)_{\lambda} + g_{\mu\lambda}(p-r)_{\nu} + g_{\nu\lambda}(r-q)_{\mu}]~,
\label{43}
\end{equation} with
\begin{equation} g_{W^-W^+\gamma} = g~\sin \theta_W = e \quad {\rm and} \quad g_{W^-W^+Z} = g~\cos \theta_W~.
\label{44}
\end{equation} This form of the triple gauge vertex is very special: in general, there could be departures from the above SM
expression, even restricting us to $SU(2)\otimes U(1)$ gauge symmetric and C and P invariant couplings. In fact some small
corrections are already induced by the radiative corrections. But, in principle, more important could be the modifications
induced by some new physics effect. The experimental testing of the triple gauge vertices is presently underway at LEP2 and
limits on departures from the SM couplings have also been obtained at the Tevatron and elsewhere.

\begin{figure}
\hglue4.0cm
\epsfig{figure=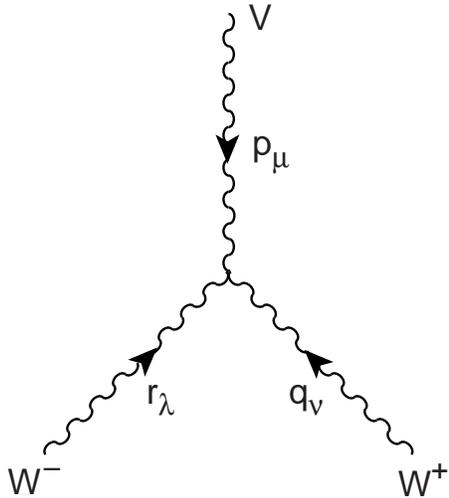,width=6cm}
\caption[ ]{The three-gauge boson vertex: $V=\gamma,Z$}
\end{figure} 

We now turn to the Higgs sector of the EW Lagrangian. Here we simply review the formalism of the
Higgs mechanism applied to the EW theory. In the next section we shall make a more general and detailed
discussion of the physics of the EW symmetry breaking. The Higgs Lagrangian is specified by the gauge
principle and the requirement of renormalizability to be
\begin{equation} {\cal L}_{\rm Higgs} = (D_{\mu}\phi)^{\dag}(D^{\mu}\phi) - V(\phi^{\dag}\phi) -
\bar \psi_L \Gamma \psi_R \phi - \bar \psi_R \Gamma^{\dag} \psi_L \phi^{\dag}~,
\label{45}
\end{equation} where $\phi$ is a column vector including all Higgs fields; it transforms as a reducible representation
of the gauge group. The quantities $\Gamma$ (which include all coupling constants) are matrices that make the Yukawa
couplings invariant under the Lorentz and gauge groups. The potential $V(\phi^{\dag}\phi)$, symmetric under $SU(2)
\otimes  U(1)$, contains, at most, quartic terms in $\phi$ so that the theory is renormalizable:
\beq
V(\phi^{\dag}\phi)=-\frac{1}{2}\mu^2\phi^{\dag}\phi+\frac{1}{4}\lambda(\phi^{\dag}\phi)^2\label{44a}
\eeq

As discussed in the next section, spontaneous symmetry
breaking is induced if the minimum of V Ñ which is the classical analogue of the quantum mechanical vacuum state (both
are the states of minimum energy) Ñ is obtained for non-vanishing $\phi$ values. Precisely, we denote the vacuum
expectation value (VEV) of $\phi$, i.e. the position of the minimum, by $v$:
\begin{equation}
\langle 0 |\phi (x)|0 \rangle = v \not= 0~.
\label{46}
\end{equation}

The fermion mass matrix is obtained from the Yukawa couplings by replacing $\phi (x)$ by $v$:
\begin{equation} M = \bar \psi_L~{\cal M} \psi_R + \bar \psi_R {\cal M}^{\dag}\psi_L~,
\label{47}
\end{equation} with
\begin{equation} {\cal M} = \Gamma \cdot v~.
\label{48}
\end{equation} In the MSM, where all left fermions $\psi_L$ are doublets and all right fermions $\psi_R$ are singlets,
only Higgs doublets can contribute to fermion masses. There are enough free couplings in $\Gamma$, so that one single
complex Higgs doublet is indeed sufficient to generate the most general fermion mass matrix. It is important to observe
that by a suitable change of basis we can always make the matrix ${\cal M}$ Hermitian, $\gamma_5$-free, and diagonal. In
fact, we can make separate unitary transformations on $\psi_L$ and $\psi_R$ according to
\begin{equation}
\psi'_L = U\psi_L, \quad \psi'_R = V\psi_R
\label{49}
\end{equation} and consequently
\begin{equation} {\cal M} \rightarrow {\cal M}' = U^{\dag}{\cal M}V~.
\label{50}
\end{equation} This transformation does not alter the general structure of the fermion couplings in ${\cal L}_{\rm
symm}$.

 If only one Higgs doublet is present, the change of basis that makes ${\cal M}$ diagonal will at the same time
diagonalize also the fermion--Higgs Yukawa couplings. Thus, in this case, no flavour-changing neutral Higgs exchanges
are present. This is not true, in general, when there are several Higgs doublets. But one Higgs doublet for each
electric charge sector i.e. one doublet coupled only to $u$-type quarks, one doublet to $d$-type quarks, one doublet to
charged leptons would also be all right, because the mass matrices of fermions with different charges are
diagonalized separately. For several Higgs doublets in a given charge sector it is also possible to generate CP
violation by complex phases in the Higgs couplings. In the presence of six quark flavours, this CP-violation mechanism is
not necessary. In fact, at the moment, the simplest model with only one Higgs doublet seems adequate for describing all
observed phenomena.

We now consider the gauge-boson masses and their couplings to the Higgs. These effects are induced by the
$(D_{\mu}\phi)^{\dag}(D^{\mu}\phi)$ term in
${\cal L}_{\rm Higgs}$ [Eq. (\ref{45})], where
\begin{equation} D_{\mu}\phi = \left[ \partial_{\mu} + ig \sum^3_{A=1} t^AW^A_{\mu} + ig'(Y/2)B_{\mu} \right] \phi~.
\label{51}
\end{equation} Here $t^A$ and $1/2Y$ are the $SU(2) \otimes U(1)$ generators in the reducible representation spanned by
$\phi$. Not only doublets but all non-singlet Higgs representations can contribute to gauge-boson masses. The condition
that the photon remains massless is equivalent to the condition that the vacuum is electrically neutral:
\begin{equation} Q|v\rangle = (t^3 + \frac{1}{2}Y)|v \rangle = 0~.
\label{52}
\end{equation} The charged $W$ mass is given by the quadratic terms in the $W$ field arising from
${\cal L}_{\rm Higgs}$, when $\phi (x)$ is replaced by $v$. We obtain
\begin{equation} m^2_WW^+_{\mu}W^{- \mu} = g^2|(t^+v/ \sqrt 2)|^2 W^+_{\mu}W^{- \mu}~,
\label{53}
\end{equation} whilst for the $Z$ mass we get [recalling Eq. (\ref{32})]
\begin{equation}
\frac{1}{2}m^2_ZZ_{\mu}Z^{\mu} = |[g \cos \theta_Wt^3 - g' \sin
\theta_W(Y/2)]v|^2Z_{\mu}Z^{\mu}~,
\label{54}
\end{equation} where the factor of 1/2 on the left-hand side is the correct normalization for the definition of the
mass of a neutral field. By using Eq. (\ref{52}), relating the action of $t^3$ and $1/2Y$ on the vacuum $v$, and Eqs.
(\ref{34}), we obtain
\begin{equation}
\frac{1}{2}m^2_Z = (g \cos \theta_W + g' \sin \theta_W)^2 |t^3v|^2 = (g^2/ \cos^2 \theta_W)|t^3v|^2~.
\label{55}
\end{equation} For Higgs doublets
\begin{equation}
\phi = \pmatrix { \phi^+ \cr
\phi^0}, \quad v = \pmatrix{ 0 \cr v}~, 
\label{56}
\end{equation} we have
\begin{equation} |t^+v|^2 = v^2, \quad |t^3v|^2 = 1/4v^2~,
\label{57}
\end{equation} so that
\begin{equation} m^2_W = 1/2g^2v^2, \quad m^2_Z = 1/2g^2v^2/\cos^2\theta_W~.
\label{58}
\end{equation} Note that by using Eq. (\ref{37}) we obtain
\begin{equation} v = 2^{-3/4}G^{-1/2}_F = 174.1~{\rm GeV}~.
\label{59}
\end{equation} It is also evident that for Higgs doublets
\begin{equation}
\rho_0 = m^2_W/m^2_Z \cos^2\theta_W = 1~.
\label{60}
\end{equation}

This relation is typical of one or more Higgs doublets and would be spoiled by the existence of Higgs triplets etc. In
general,
\begin{equation}
\rho_0 = \sum_i((t_i)^2 - (t^3_i)^2 + t_i ) v^2_i/ \sum _i2(t^3_i)^2v^2_i
\label{61}
\end{equation} for several Higgses with VEVs $v_i$, weak isospin $t_i$, and $z$-component $t^3_i$. These results are
valid at the tree level and are modified by calculable EW radiative corrections, as discussed in Section 6.

In the minimal version of the SM only one Higgs doublet is present. Then the fermion--Higgs couplings are in proportion to
the fermion masses. In fact, from the Yukawa couplings $g_{\phi
\bar f f}(\bar f_L \phi f_R + h.c.)$, the mass $m_f$ is obtained by replacing
$\phi$ by $v$, so that $ m_f = g_{\phi \bar f f} v $. In the minimal SM
three out of the four Hermitian fields are removed from the physical spectrum by
the Higgs mechanism and become the longitudinal modes of $W^+, W^-$, and $Z$. The fourth neutral Higgs is physical and
should be found. If more doublets are present, two more charged and two more neutral Higgs scalars should be around for
each additional doublet.

The couplings of the physical Higgs $H$ to the gauge bosons can be simply obtained from ${\cal L}_{\rm
Higgs}$, by the replacement
\begin{equation}
\phi(x) = \pmatrix{ \phi^+(x) \cr
\phi^0(x)} \rightarrow 
\pmatrix {0 \cr v + (H/\sqrt2)}~,
\label{62}
\end{equation} [so that $(D_{\mu}\phi)^{\dag}(D^{\mu}\phi) = 1/2(\partial_{\mu}H)^2 + ...]$, with the result
\begin{eqnarray} {\cal L} [H,W,Z] &=& g^2(v/\sqrt 2)W^+_{\mu}W^{-\mu} H + (g^2 /4)W^+_{\mu}W^{-\mu}H^2 \nonumber \\ &&
+ [(g^2vZ_{\mu}Z^{\mu})/(2 \sqrt 2 \cos^2\theta_W)]H \nonumber \\ &&+ [g^2/(8
\cos^2\theta_W)]Z_{\mu}Z^{\mu}H^2~.
\label{63}
\end{eqnarray}

In the minimal SM the Higgs mass $m^2_H\sim \lambda v^2$ is of order of the weak scale v. We will discuss in sect.9 the
direct experimental limit on $m_H$ from LEP, which is $m_H\gappeq 113~GeV$. We shall also see in sect.9 , that, if there
is no physics beyond the SM up to a large scale $\Lambda$, then, on theoretical grounds, $m_H$ can only be within a narrow
range between 135 and 180 GeV. But the interval is enlarged if there is new physics nearby. Also the lower limit depends
critically on the assumption of only one doublet. The dominant decay mode of the Higgs is in the $b \bar b$ channel below
the WW threshold, while the $W^+W^-$ channel is dominant for sufficiently large $m_H$. The width is small below the WW
threshold, not exceeding a few MeV, but increases steeply beyond the threshold, reaching the asymptotic value of $\Gamma\sim
1/2 m^3_H$ at large $m_H$, where all energies are in TeV.

\section{The Higgs Mechanism}

The gauge symmetry of the Standard Model was difficult to discover because it is well hidden in nature. The only
observed gauge boson that is massless is the photon. The gluons are presumed massless but are unobservable because of
confinement, and the $W$ and $Z$ weak bosons carry a heavy mass. Actually a major difficulty in unifying weak and
electromagnetic interactions was the fact that e.m. interactions have infinite range $(m_{\gamma} = 0)$, whilst the weak
forces have a very short range, owing to
$m_{W,Z} \not= 0$.

The solution of this problem is in the concept of spontaneous symmetry breaking, which was borrowed from statistical
mechanics. 

Consider a ferromagnet at zero magnetic field in the Landau--Ginzburg approximation. The free energy in terms of the
temperature $T$ and the magnetization {\bf M} can be written as
\begin{equation} F({\bf M}, T) \simeq F_0(T) + 1/2~\mu^2(T){\bf M}^2 + 1/4~\lambda(T)({\bf M}^2)^2 + ...~.
\label{64}
\end{equation} This is an expansion which is valid at small magnetization.  The neglect of terms of higher order in
$\vec M^2$ is the analogue in this context of the renormalizability criterion. Also, $\lambda(T) > 0$ is assumed for
stability; $F$ is invariant under rotations, i.e. all directions of {\bf M} in space are equivalent. The minimum
condition for $F$ reads
\begin{equation}
\partial F/\partial M = 0, \quad [\mu^2(T) + \lambda(T){\bf M}^2]{\bf M} = 0~.
\label{65}
\end{equation} There are two cases. If $\mu^2 > 0$, then the only solution is ${\bf M} = 0$, there is no magnetization,
and the rotation symmetry is respected. If $\mu^2 < 0$, then another solution appears, which is
\begin{equation} |{\bf M}_0|^2 = -\mu^2/\lambda~.
\label{66}
\end{equation} The direction chosen by the vector ${\bf M}_0$ is a breaking of the rotation symmetry. The critical
temperature $T_{\rm crit}$ is where $\mu^2(T)$ changes sign:
\begin{equation}
\mu^2(T_{\rm crit}) = 0~.
\label{67}
\end{equation} It is simple to realize that the Goldstone theorem holds. It states that when spontaneous symmetry
breaking takes place, there is always a zero-mass mode in the spectrum. In a classical context this can be proven as
follows. Consider a Lagrangian
\begin{equation} {\cal L} = |\partial_{\mu}\phi|^2 - V(\phi)
\label{68}
\end{equation} symmetric under the infinitesimal transformations
\begin{equation}
\phi \rightarrow  \phi' = \phi + \delta \phi, \quad
\delta \phi_i = i \delta \theta t_{ij}\phi_j~.
\label{69}
\end{equation} The minimum condition on $V$ that identifies the equilibrium position (or the ground state in quantum
language) is
\begin{equation} (\partial V/\partial \phi_i)(\phi_i = \phi^0_i) = 0~.
\label{70}
\end{equation} The symmetry of $V$ implies that
\begin{equation}
\delta V = (\partial V/\partial \phi_i)\delta \phi_i = i \delta \theta(\partial V/\partial \phi_i)t_{ij}\phi_j = 0~.
\label{71}
\end{equation} By taking a second derivative at the minimum $\phi_i = \phi^0_i$ of the previous equation, we obtain
\begin{equation}
\partial^2V/\partial \phi_k\partial \phi_i (\phi_i =
\phi^0_i)t_{ij}\phi^0_i + \frac{\partial V}{\partial \phi_i} (\phi_i =
\phi^0_i)t_{ik} = 0~.
\label{72}
\end{equation} The second term vanishes owing to the minimum condition, Eq. (\ref{70}). We then find
\begin{equation}
\partial^2V/\partial \phi_k\partial \phi_i ~(\phi_i = \phi^0_i)t_{ij}\phi^0_j = 0~.
\label{73}
\end{equation} The second derivatives $M^2_{ki} = (\partial^2V/\partial \phi_k \partial
\phi_i)(\phi_i = \phi^0_i)$ define the squared mass matrix. Thus the above equation in matrix notation can be read as
\begin{equation} M^2 t\phi^0 = 0~,
\label{74}
\end{equation} which shows that if the vector $(t\phi^0)$ is non-vanishing, i.e. there is some generator that shifts
the ground state into some other state with the same energy, then $t \phi^0$ is an eigenstate of the squared mass
matrix with zero eigenvalue. Therefore, a massless mode is associated with each broken generator.

When spontaneous symmetry breaking takes place in a gauge theory, the massless Goldstone mode exists, but it is
unphysical and disappears from the spectrum. It becomes, in fact, the third helicity state of a gauge boson that takes
mass. This is the Higgs mechanism. Consider, for example, the simplest Higgs model described by the Lagrangian
\begin{equation} {\cal L} = -\frac{1}{4}~F^2_{\mu\nu} + |(\partial_{\mu} - ieA_{\mu})\phi|^2 +
\frac{1}{2} \mu^2 \phi^*\phi - (\lambda/4)(\phi^*\phi)^2~.
\label{75}
\end{equation} Note the `wrong' sign in front of the mass term for the scalar field $\phi$, which is necessary for the
spontaneous symmetry breaking to take place. The above Lagrangian is invariant under the $U(1)$ gauge symmetry
\begin{equation} A_{\mu} \rightarrow A'_{\mu} = A_{\mu} - (1/e)\partial_{\mu}\theta(x), \quad
\phi \rightarrow \phi' = \phi ~{\rm exp}[i\theta(x)]~.
\label{76}
\end{equation} Let $\phi^0 = v \not= 0$, with $v$ real, be the ground state that minimizes the potential and induces
the spontaneous symmetry breaking. Making use of gauge invariance, we can make the change of variables
\begin{eqnarray} &&\phi(x) \rightarrow (1/\sqrt 2)[\rho(x) + v]~{\rm exp}[i \zeta(x)/v]~, \nonumber \\ &&A_{\mu}(x)
\rightarrow A_{\mu} - (1/ev)\partial_{\mu}  \zeta(x).
\label{77}
\end{eqnarray} Then $\rho = 0$ is the position of the minimum, and the Lagrangian becomes
\begin{equation} {\cal L} = -\frac{1}{4}F^2_{\mu\nu} + \frac{1}{2}e^2v^2A^2_{\mu} + \frac{1}{2} e^2\rho^2A^2_{\mu} +
e^2\rho vA^2_{\mu} + {\cal L}(\rho)~.
\label{78}
\end{equation} The field $\zeta(x)$, which corresponds to the would-be Goldstone boson, disappears, whilst the mass
term $\frac{1}{2}e^2v^2A^2_{\mu}$ for $A_{\mu}$ is now present; $\rho$ is the massive Higgs particle.

The Higgs mechanism is realized in well-known physical situations. For a superconductor in the Landau--Ginzburg
approximation the free energy can be written as
\begin{equation} F = F_0 + \frac{1}{2}{\bf B}^2 + |({\bf \nabla} - 2ie{\bf A})\phi|^2/4m -
\alpha|\phi|^2 + \beta|\phi|^4~.
\label{79}
\end{equation} 

Here {\bf B} is the magnetic field, $|\phi|^2$ is the Cooper pair $(e^-e^-)$ density, 2$e$ and 2$m$ are the charge and
mass of the Cooper pair. The 'wrong' sign of $\alpha$ leads to $\phi \not= 0$ at the minimum. This is precisely the
non-relativistic analogue of the Higgs model of the previous example. The Higgs mechanism implies the absence of
propagation of massless phonons (states with dispersion relation ~$\omega = kv$ with constant $v$). Also the mass term
for {\bf A} is manifested by the exponential decrease of {\bf B} inside the superconductor (Meissner effect).

\section{The CKM Matrix}

Weak charged currents are the only tree level interactions in the SM that change flavour: by emission of a W an
up-type quark is turned into a  down-type quark, or a $\nu_l$ neutrino is turned into a $l^-$
charged lepton (all fermions are letf-handed). If we start from an up quark that is a mass
eigenstate, emission of a W turns it into a down-type quark state d' (the weak isospin partner of
u) that in general is not a mass eigenstate. In general, the mass eigenstates and the weak
eigenstates do not coincide and a unitary transformation connects the two sets:
\beq
\left(\matrix{d^\prime\cr s^\prime\cr b^\prime}\right)=V\left(\matrix{d\cr s\cr b}\right)\label{km1}
\eeq
V is the Cabibbo-Kobayashi-Maskawa matrix.
Thus in terms of mass eigenstates the charged weak current of quarks is of the form:
\beq
J^+_{\mu}\propto\bar u \gamma_{\mu}(1-\gamma_5)t^+ Vd 
\label{km2}
\eeq
Since V is unitary (i.e. $VV^\dagger=V^\dagger V=1$) and commutes with $T^2$, $T_3$ and Q (because all d-type quarks
have the same isospin and charge) the neutral current couplings are diagonal both in the primed and unprimed basis (if
the Z down-type quark current is abbreviated as $\bar d^\prime \Gamma d^\prime$ then by changing basis we get $\bar d
V^\dagger \Gamma V d$ and V and $\Gamma$ commute because, as seen from eq.(\ref{41}), $\Gamma$ is made of Dirac
matrices and $T_3$ and Q generator matrices). It follows that $\bar d^\prime \Gamma d^\prime =\bar d \Gamma d$. This is
the GIM mechanism that ensures natural flavour conservation of the neutral current couplings at the tree level. 

For N generations of quarks, V is a NxN unitary matrix that depends on $N^2$ real numbers ($N^2$ complex entries with
$N^2$ unitarity constraints). However, the $2N$ phases of up- and down-type quarks are not observable. Note that an
overall phase drops away from the expression of the current in eq.(\ref{km2}), so that only $2N-1$ phases can affect V.
In total, V depends on $N^2-2N+1=(N-1)^2$ real physical parameters. A similar counting gives $N(N-1)/2$ as the number of
independent parameters in an orthogonal NxN matrix. This implies that in V we have $N(N-1)/2$ mixing angles and
$(N-1)^2-N(N-1)/2$ phases: for $N=2$ one mixing angle (the Cabibbo angle) and no phase, for $N=3$ three angles and one
phase etc. 

Given the experimental near diagonal structure of V a convenient parametrisation is the one proposed by
Maiani. One starts by the definition:
\beq
\vert d'\rangle=c_{13}\vert d_C\rangle+s_{13} e^{-i\phi}\vert b\rangle
\label{km3}
\eeq
where $c_{13}\equiv cos\theta_{13}$, $s_{13}\equiv sin\theta_{13}$ (analogous shorthand notations will be used in the
following), $d_C$ is the Cabibbo down quark and  $\theta_{12}\equiv\theta_C$ is the
Cabibbo angle (experimentally $s_{12}\equiv\lambda\sim 0.22$).
\beq
\vert d_C\rangle=c_{12}\vert d\rangle+s_{12} \vert s\rangle\label{km4}
\eeq
Note that in a four quark model the Cabibbo angle fixes both the ratio of couplings $(u\rightarrow
d)/(\nu_e\rightarrow e)$ and the ratio of $(u\rightarrow
d)/(u\rightarrow s)$. In a six quark model one has to choose which to keep as a definition of the Cabibbo angle.
Here the second definition is taken and, in fact the $u\rightarrow d$ coupling is given by $V_{ud}=c_{13} c_{12}$ so
that it is no longer specified by $\theta_{12}$ only. Also note that we can certainly fix the phases of u, d, s so
that a real coefficient appears in front of $d_C$ in eq.(\ref{km3}). We now choose a basis of two orthonormal vectors,
both orthogonal to $\vert d'\rangle$:
\beq
\vert s_C\rangle=-s_{12}\vert d\rangle+c_{12} \vert s\rangle,~~~~~~~\vert v\rangle=-s_{13} e^{i\phi}\vert
d_C\rangle+c_{13} \vert b\rangle\label{km5}
\eeq 
Here $\vert s_C\rangle$ is the Cabibbo s quark. Clearly s' and b' must be othonormal superpositions of the above base
vectors defined in terms of an angle $\theta_{23}$:
\beq
\vert s'\rangle=c_{23}\vert s_C\rangle+s_{23} \vert v\rangle,~~~~~~\vert b'\rangle=-s_{23}\vert
s_C\rangle+c_{23} \vert v\rangle\label{km6}
\eeq 
The general expression of $V_{ij}$ can be obtained from the above equations. But a considerable notational
simplification is gained if one takes into account that from experiment we know that $s_{12}\equiv\lambda$, $s_{23}\sim
o(\lambda^2)$ and 
$s_{13}\sim o(\lambda^3)$ are increasingly small and of the indicated orders of magnitude. Thus, following Wolfenstein
one can set:
\beq
s_{12}\equiv\lambda,~~~~~~~~s_{23}=A\lambda^2,~~~~~~~~s_{13}e^{-i\phi}=A\lambda^3(\rho-i\eta)\label{km7}
\eeq
As a result, by neglecting terms of higher order in $\lambda$ one can write down:
\beq
V= 
\left[\matrix{
V_{ud}&V_{us}&V_{ub} \cr
V_{cd}&V_{cs}&V_{cb}\cr
V_{td}&V_{ts}&V_{tb}     } 
\right ]~~~\sim~~~\left[\matrix{
1-\frac{\lambda^2}{2}&\lambda&A\lambda^3(\rho-i\eta) \cr
-\lambda&1-\frac{\lambda^2}{2}&A\lambda^2\cr
A\lambda^3(1-\rho-i\eta)&-A\lambda^2&1     } 
\right ].
\label{km8}
\eeq 
Indicative values of the CKM parameters as obtained from experiment are (a survey of the current status of the CKM
parameters can be found in ref.\cite{pdg}):
\bea
\lambda=0.2196\pm0.0023\nonumber\\
A=0.83\pm0.04\nonumber\\
\sqrt{\rho^2+\eta^2}=0.4\pm0.1;~~~~~ \eta\sim 0.3\pm0.1\label{km9}
\eea

In the SM the non vanishing of the $\eta$ parameter is the only source of CP violation. Unitarity of the CKM matrix V implies
relations of the form
$\sum_a V_{ba}V^*_{ca}=\delta_{bc}$. In most cases these relations do not imply particularly instructive constraints on the
Wolfenstein parameters. But when the three terms in the sum are of comparable magnitude we get interesting information. The
three numbers which must add to zero form a closed triangle in the complex plane, with sides of comparable length. This is the
case for the t-u triangle (Bjorken triangle) shown in fig.2:

\beq
V_{td}V^*_{ud}+V_{ts}V^*_{us}+V_{tb}V^*_{ub}=0\label{km10}
\eeq
All terms are of order $\lambda^3$. For $\eta$=0 the triangle would flatten down to vanishing area. In fact
the area of the triangle, J of order $J\sim \eta A^2 \lambda^6$, is the Jarlskog invariant (its value is independent of the
parametrization). In the SM all CP violating observables must be proportional to J, hence to the area of the triangle
or to $\eta$. The most direct and solid evidence for
J non vanishing is obtained from the measurement of $\epsilon$ in K decay. Additional direct evidence is being obtained
from the measurement of $\sin{2\beta}$ in B decay. 

\begin{figure}
\hglue4.0cm
\epsfig{figure=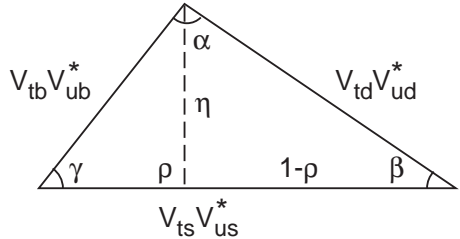,width=6cm}
\caption[ ]{The Bjorken triangle corresponding to eq.(\ref{km10}):}
\end{figure}
 
We have only discussed flavour mixing for quarks. But, clearly, if neutrino masses exist, as indicated by
neutrino oscillations (see section 8.2.3), then a similar mixing matrix must also be introduced in the leptonic sector. 

\section{Renormalisation and Higher Order Corrections}

The Higgs mechanism gives masses to the Z, the $W^\pm$ and to fermions while the Lagrangian density is still
symmetric. In particular the gauge Ward identities and the conservation of the gauge currents are preserved. The validity of
these relations is an essential ingredient for renormalisability. For example the massive gauge boson propagator would have
a bad ultraviolet behaviour:
\beq
W_{\mu\nu}=\frac{-g_{\mu\nu}+\frac{q_\mu q_\nu}{m^2_W}}{q^2-m^2_W}\label{prop}
\eeq
But if the propagator is sandwiched between conserved currents $J_\mu$ the bad terms in $q_\mu q_\nu$ give a vanishing
contribution because $q_\mu J^\mu=0$ and the high energy behaviour is like for a scalar particle and compatible with
renormalisation. 

The fundamental theorem that in general a gauge theory with spontaneous symmetry breaking and the Higgs
mechanism is renormalisable was proven by 't Hooft. For a chiral theory like the SM an additional complication arises from
the existence of chiral anomalies. But this problem is avoided in the SM because the quantum numbers of the quarks and
leptons in each generation imply a remarkable (and apparently miracoulous) cancellation of the anomaly, as originally
observed by Bouchiat, Iliopoulos and Meyer. In quantum field theory one encounters an anomaly when a symmetry of the
classical lagrangian is broken by the process of quantisation, regularisation and renormalisation of the theory. For
example, in massless QCD there is no mass scale in the classical lagrangian. Thus one would predict that dimensionless
quantities in processes with only one large energy scale Q cannot depend on Q and must be constants. As well known this
naive statement is false. The process of regularisation and renormalisation necessarily introduces an energy scale which is
essentially the scale where renormalised quantities are defined. For example the renormalised coupling must be defined from
the vertices at some scale. This scale
$\mu$ cannot be zero because of infrared divergences. The scale $\mu$ destroys scale invariance because dimensionless
quantities can now depend on $Q/\mu$. The famous $\Lambda_{QCD}$ parameter is a tradeoff of $\mu$ and leads to scale
invariance breaking. Of direct relevance for the EW theory is the Adler-Bell-Jackiw chiral anomaly. The classical lagrangian
of a theory with massless fermions is invariant under a U(1) chiral transformations
$\psi\prime=e^{i\gamma_5\theta}\psi$. The associated axial Noether current is conserved at the classical level. But, at the
quantum level, chiral symmetry is broken due to the ABJ anomaly and the current is not conserved. The chiral breaking is
introduced by a clash between chiral symmetry, gauge invariance and the regularisation procedure. The anomaly is generated
by triangular fermion loops with one axial and two vector vertices (fig.3).
  For neutral currents (Z and
$\gamma$) the axial coupling is proportional to the 3rd component of weak isospin $t_3$, while vector couplings are
proportional to a linear combination of
$t_3$ and the electric charge Q. Thus in order for the chiral anomaly to vanish all traces of the form $tr\{t_3QQ\}$,
$tr\{t_3t_3Q\}$, $tr\{t_3t_3t_3\}$ (and also  $tr\{t_+t_-t_3\}$ when charged currents are also included) must vanish, where
the trace is extended over all fermions in the theory that can circulate in the loop. Now all these
traces happen to vanish for each fermion family separately. For example take $tr\{t_3QQ\}$. In one family there are, with
$t_3=+1/2$, three colours of up quarks with charge $Q=+2/3$ and one neutrino with $Q=0$ and, with $t_3=-1/2$, three colours
of down quarks with charge $Q=-1/3$ and one $l^-$ with $Q=-1$. Thus we obtain $tr\{t_3QQ\}=1/2~3~4/9-1/2~3~1/9-1/2~1=0$.
This impressive cancellation suggests an interplay among weak isospin, charge and colour quantum numbers which appears as a
miracle from the point of view of the low energy theory but is more understandable from the point of view of the high energy
theory. For example in GUTs there are similar relations where charge quantisation and colour are related: in the 5 of SU(5)
we have the content $(d,d,d,e^+,\bar\nu)$ and the charge generator has a vanishing trace in each SU(5) representation (the
condition of unit determinant, represented by the letter S in the SU(5) group name, translates into zero trace for the
generators). Thus the charge of d quarks is -1/3 of the positron charge because there are three colours.

\begin{figure}
\hglue4.0cm
\epsfig{figure=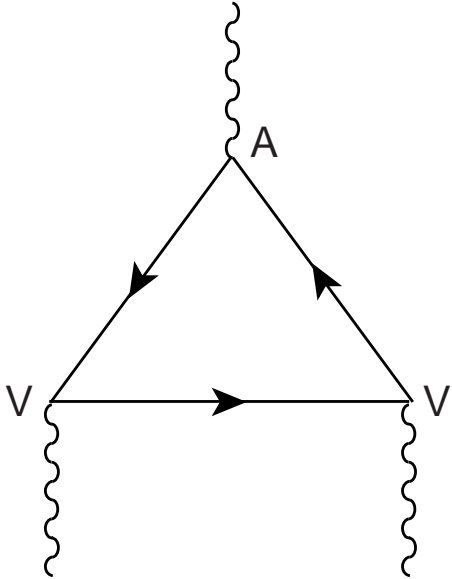,width=6cm}
\caption[ ]{Triangle diagram that generates the ABJ anomaly.}
\end{figure}

Since the SM theory is renormalisable higher order perturbative corrections can be reliably computed. Radiative corrections
are very important for precision EW tests. The SM inherits all successes of the old V-A theory of charged currents and of
QED. Modern tests focus on neutral current processes, the W mass and the measurement of triple gauge vertices. For Z physics
and the W mass the state of the art computation of radiative corrections include the complete one loop diagrams and selected
dominant two loop corrections. In addition some resummation techniques are also implemented, like Dyson resummation of vacuum
polarisation functions and important renormalisation group improvements for large QED and QCD logarithms. We now discuss in
more detail sets of large radiative corrections which are particularly significant \cite{radcorr}.

A set of important quantitative contributions to the radiative corrections arise from large logarithms [e.g. terms of the
form $(\alpha/\pi ~{\rm ln}~(m_Z/m_{f_\ell}))^n$ where $f_{\ell}$ is a light fermion]. The sequences of leading and
close-to-leading logarithms are fixed by well-known and consolidated techniques ($\beta$ functions, anomalous
dimensions, penguin-like diagrams, etc.). For example, large logarithms dominate the running of
$\alpha$ from $m_e$, the electron mass, up to $m_Z$. Similarly large logarithms of the form $[\alpha/\pi~{\rm
ln}~(m_Z/\mu)]^n$ also enter, for example, in the relation between $\sin^2\theta_W$ at the scales $m_Z$ (LEP, SLC) and $\mu$
(e.g. the scale of low-energy neutral-current experiments). Also, large logs from initial state radiation dramatically distort
the line shape of the Z resonance as observed at LEP1 and SLC and must be accurately taken into account in the measure of
the Z mass and total width.

For example, a considerable amount of work has deservedly been devoted to the theoretical study of the $Z$ line-shape. The
experimental accuracy on $m_Z$ obtained at LEP1 is $\delta m_Z = \pm 2.1$~MeV. This small
error  was obtained by a precise calibration of the LEP energy scale achieved by taking advantage of the transverse
polarization of the beams and implementing a sophisticated resonant spin depolarization method. Similarly, a
measurement of the total width to an accuracy $\delta \Gamma = \pm 2.4$~MeV has been achieved. The prediction of the
Z line-shape in the SM to such an accuracy has posed a formidable challenge to theory, which has been
successfully met. For the inclusive process $e^+e^- \rightarrow f \bar fX$, with $f \not= e$ (for simplicity, we leave
Bhabha scattering aside) and $X$ including $\gamma$'s and gluons, the physical cross-section can be written in the form of a
convolution
\cite{radcorr}: 
\begin{equation} \sigma(s) = \int^1_{z_0} dz~\hat \sigma(zs)G(z,s)~,
 \label{130}
\end{equation}  where $\hat \sigma$ is the reduced cross-section, and $G(z,s)$ is the radiator function that describes the
effect of initial-state radiation; $\hat \sigma$ includes the purely weak corrections, the effect of final-state radiation
(of both $\gamma$'s and gluons), and also non-factorizable terms (initial- and final-state radiation interferences, boxes,
etc.) which, being small, can be treated in lowest order and effectively absorbed in a modified $\hat \sigma$. The radiator
$G(z,s)$ has an expansion of the form
\begin{eqnarray} G(z,s) & = &
\delta(1-z) + \alpha /\pi(a_{11}L + a_{10}) + (\alpha/\pi)^2 (a_{22}L^2 + a_{11}L + a_{20}) \nonumber \\ && +~... +
(\alpha/\pi)^n~\sum^n_{i=0} a_{ni}L^i~,
\label{131}
\end{eqnarray} where $L = {\rm ln}~s/m^2_e \simeq 24.2$ for $\sqrt s \simeq m_Z$. All first- and second-order terms are known
exactly. The sequence of leading and next-to-leading logs can be exponentiated (closely following the formalism
of structure functions in QCD). For $m_Z \approx 91$~GeV, the convolution displaces the peak by  +110~MeV, and reduces it by
a factor of about 0.74. The exponentiation is important in that it amounts to a shift of about 14~MeV in the peak position.
 
Among the one loop EW radiative corrections, a very remarkable class of contributions are those terms that increase
quadratically with the top mass.  The sensitivity
of radiative corrections to $m_t$ arises from the existence of these terms. The quadratic dependence on
$m_t$ (and on other possible widely broken isospin multiplets from new physics) arises because, in spontaneously broken
gauge theories, heavy loops do not decouple. On the contrary, in QED or QCD, the running of
$\alpha$ and $\alpha_s$ at a scale $Q$ is not affected by heavy quarks with mass
$M \gg Q$. According to an intuitive decoupling theorem, diagrams with heavy virtual particles of mass $M$ can be
ignored at $Q \ll M$ provided that the couplings do not grow with $M$ and that the theory with no heavy particles is still
renormalizable. In the spontaneously broken EW gauge theories both requirements are violated. First, one important difference
with respect to unbroken gauge theories is in the longitudinal modes of weak gauge bosons. These modes are generated by the
Higgs mechanism, and their couplings grow with masses (as is also the case for the physical Higgs couplings). Second the
theory without the top quark is no more renormalisable because the gauge symmetry is broken because the doublet (t,b) would
not be complete (also the chiral anomaly would not be completely cancelled). With the observed value of
$m_t$ the quantitative importance of the terms of order $G_Fm^2_t/4\pi^2\sqrt{2}$ is substancial but not dominant (they are
enhanced by a factor $m^2_t/m^2_W\sim 5$ with respect to ordinary terms). Both the large logarithms and the
$G_Fm^2_t$ terms have a simple structure and are to a large extent universal, i.e. common to a wide class of processes. In
particular the $G_Fm^2_t$ terms appear in vacuum polarization diagrams which are universal and in the $Z\rightarrow b \bar b$
vertex which is not (this vertex is connected with the top quark which runs in the loop, while other types of heavy particles
could  in principle also contribute to vacuum polarization diagrams). Their study is important for an understanding of the
pattern of radiative corrections. One can also derive approximate formulae (e.g. improved Born approximations), which can be
useful in cases where a limited precision may be adequate.  More in general, another very important consequence of non
decoupling is that precision tests of the electroweak theory may be sensitive to new physics even if the new particles are
too heavy for their direct production.

While radiative corrections are quite sensitive to the top mass, they are unfortunately much less dependent on the Higgs
mass. If they were sufficiently sensitive by now we would precisely know the mass of the Higgs. But the dependence of one
loop diagrams on
$m_H$ is only logarithmic:
$\sim G_Fm^2_W
\log(m^2_H/m^2_W)$. Quadratic terms $\sim G^2_Fm^2_H$ only appear at two loops and are too small to be important. The
difference with the top case is that the difference $m^2_t-m^2_b$ is a direct breaking of the gauge symmetry that
already affects the one loop corrections, while the Higgs couplings are "custodial" SU(2) symmetric in lowest order.

The basic tree level relations:
\beq
\frac{g^2}{8m^2_W}=\frac{G_F}{\sqrt{2}},~~~~~~g^2\sin^2\theta_W=e^2=4\pi\alpha\label{bb1}
\eeq
can be combined into
\beq
\sin^2\theta_W=\frac{\pi\alpha}{\sqrt{2}G_Fm^2_W}\label{bb2}
\eeq
A different definition of $\sin^2\theta_W$ is from the gauge boson masses:
\beq
\frac{m^2_W}{m^2_Z\cos^2\theta_W}=\rho_0=1~~~\Longrightarrow~~~\sin^2\theta_W=1-\frac{m^2_W}{m^2_Z}\label{bb3}
\eeq
where $\rho_0=1$ assuming that there are only Higgs doublets. The last two relations can be put into the convenient form
\beq
(1-\frac{m^2_W}{m^2_Z})\frac{m^2_W}{m^2_Z}=\frac{\pi\alpha}{\sqrt{2}G_Fm^2_Z}\label{bb4}
\eeq
These relations are modified by radiative corrections:
\bea
(1-\frac{m^2_W}{m^2_Z})\frac{m^2_W}{m^2_Z}=\frac{\pi\alpha(m_Z)}{\sqrt{2}G_Fm^2_Z}\frac{1}{1-\Delta r_W}\nonumber\\
\frac{m^2_W}{m^2_Z\cos^2\theta_W}=1+\rho_m\label{bb5}
\eea
In the first relation the replacement of $\alpha$ with the running coupling at the Z mass $\alpha(m_Z)$ makes $\Delta r_W$
completely determined by the purely weak corrections. This relation defines $\Delta r_W$ unambigously, once the meaning of
$\alpha(m_Z)$ is specified. On the contrary, in the second relation $\Delta \rho_m$ depends on the definition of
$\sin^2\theta_W$ beyond the tree level. For LEP physics $\sin^2\theta_W$ is usually defined from the
$Z\rightarrow\mu^+\mu^-$ effective vertex. At the tree level we have:
\beq
Z\rightarrow f^+f^-=\frac{g}{2\cos\theta_W}\bar f\gamma_\mu(g^f_V-g^f_A\gamma_5)f \label{xyz}
\eeq
with $g^{f2}_A=1/4$ and $g^f_V/g^f_A=1-4|Q_f|\sin^2\theta_W$. Beyond the tree level a corrected vertex can be written down
in the same form of eq.(\ref{xyz}) in terms of modified effective couplings. Then $\sin^2\theta_W\equiv\sin^2\theta_{eff}$
is in general defined through the muon vertex:
\bea
g^\mu_V/g^\mu_A&=&1-4\sin^2\theta_{eff}\nonumber\\
\sin^2\theta_{eff}&=&(1+\Delta k)s^2_0,~~~~~~~s^2_0 c^2_0=\frac{\pi\alpha(m_Z)}{\sqrt{2}G_Fm^2_Z}\nonumber\\
g^{\mu2}_A&=&\frac{1}{4}(1+\Delta\rho)\label{bb7}
\eea
Actually, since in the SM lepton universality is only broken by masses and is in agreement with experiment within the
present accuracy, in practice the muon channel is replaced with the average over charged leptons.

We end this discussion by
writing a symbolic equation that summarises the status of what has been computed up to now for the radiative corrections
$\Delta r_W$, $\Delta \rho$ and $\Delta k$:
\beq
\Delta r_W, \Delta \rho, \Delta k=g^2 \frac{m^2_t}{m^2_W}(1+\alpha_s+\alpha^2_s) +g^2(1+\alpha_s+\sim\alpha^2_s) + g^4
\frac{m^4_t}{m^4_W} + g^4\frac{m^2_t}{m^2_W} +...\label{bb8}
\eeq
The meaning of this relation is that the one loop terms of order $g^2$ are completely known, together with their first 
order QCD corrections (the second order QCD corrections are only estimated for the $g^2$ terms not enhanced by
$m^2_t/m^2_W$), and the terms of order $g^4$ enhanced by the ratios $m^4_t/m^4_W$ or $m^2_t/m^2_W$ are also known.

In recent years new powerful tests of the SM have been performed mainly at LEP but also
at SLC and at the Tevatron. The running of LEP1 was terminated in 1995 and close-to-final results of the data
analysis are now available. The SLC is also finished. The experiments at the Z resonance have enormously
improved the accuracy in the electroweak neutral current sector. The top quark has been at last
found at the Tevatron and the errors on $m_Z$ and $\sin^2\theta_{eff}$ went down by two and one orders of magnitude
respectively since the start of LEP in 1989. The LEP2 programme is almost completed by now. The validity
of the SM has been confirmed to a level that we can say was unexpected at the beginning. In the present data
there is no significant evidence for departures from the SM, no convincing hint of new physics. The impressive success of the
SM poses strong limitations on the possible forms of  new physics. Favoured are models of the Higgs sector and of new physics
that preserve the SM structure  and only very delicately improve it, as is the case for fundamental Higgs(es) and
Supersymmetry. Disfavoured are models with a nearby strong non perturbative regime that  almost inevitably
would affect the radiative corrections, as for composite Higgs(es) or technicolor and its variants. 

\section{Why we do Believe in the SM: Precision Tests}
\subsection{ Precision Electroweak Data and the Standard Model}

The relevant electro-weak data together with their SM values are presented in table 1 \cite{ew}.  The
SM predictions correspond to a fit of all the available data (including the directly measured values of $m_t$
and
$m_W$) in terms of $m_t$, $m_H$ and $\alpha_s(m_Z)$, described later in sect., table 4.

Other important derived quantities are, for example, $N_\nu$ the number of light neutrinos,
obtained from the invisible width: $N_\nu=2.9835(83)$, which is $2\sigma$ below 3 and indicates that only three fermion
generations exist with $m_\nu <45~GeV$. This is one of the most important results of LEP. Other important quantities are the
leptonic width
$\Gamma_l$, averaged over e,
$\mu$ and
$\tau$:
$\Gamma_l= 83.959(89) MeV$ and the hadronic width $\Gamma_h= 1743.9(2.0) MeV$.   

For indicative purposes, in table  the "pulls" are also shown, defined as: pull = (data point- fit
value)/(error on data point). 
At a glance we see that the agreement with the SM is quite good. The distribution of the
pulls is statistically normal. The presence of a few $\sim2\sigma$ deviations is what is to be expected.
For example, the atomic parity violation in Cs, a low energy experiment, shows a $2.5\sigma$ deviation. While there could be
new physics terms that only sizeably contribute to this channel (a specific contact term, a Z' unmixed with the Z), the
apparent deviation may simply be due to the difficulty of the measurement and the complicacies of the Cesium
wave-function\footnote{In a very recent paper \cite{der} new terms from the Breit interaction in the atomic-structure
calculation are shown to bring the discrepancy down to the 1$\sigma$ level.  So this problem is probably resolved.}.
One unpleasant feature of the data is the difference between the values of
$\sin^2\theta_{eff}$ measured at LEP and at SLC. The value of
$\sin^2\theta_{eff}$ is obtained from a set of combined asymmetries. From asymmetries one derives the ratio $x=g_V^l/g_A^l$
of the vector and axial vector couplings of the Z, averaged over the charged leptons. In turn $\sin^2\theta_{eff}$ is
defined by $x=1-4\sin^2\theta_{eff}$. SLD obtains x from the single measurement of
$A_{LR}$, the left-right asymmetry, which requires longitudinally polarized beams. The LEP average,
$\sin^2\theta_{eff}=0.23192(23)$, differs by
$2.2\sigma$ from the SLD value
$\sin^2\theta_{eff}=0.23099(26)$. The most
precise individual measurement at LEP is from $A^{FB}_b$: the combined LEP error on this quantity is comparable to the SLD
error, but the two values are $2.7\sigma$'s away. It is difficult to find a simple explanation for the SLD-LEP discrepancy on
$\sin^2\theta_{eff}$. In the following we will tentatively use the official average
\beq
\sin^2\theta_{eff}=0.23151\pm0.00017 \label{ew8}
\eeq	
obtained by a simple combination of the LEP-SLC data. However, one could be more conservative and enlarge the error because of
the larger dispersion.


\begin{table} 
\caption{Data on precision electroweak test}
\label{tab1}
\vglue.3cm
\begin{center}
\begin{tabular}{|l|l|l|}
\hline Quantity&Data (March 2000)       & Pull\\
\hline
$m_Z$ (GeV)&91.1871(21)  &$~~0.1$\\
$\Gamma_Z$ (GeV)        &2.4944(24)  & $-0.6 $\\
$\sigma_h$ (nb) &41.544(37)     & $~~1.7$\\
$R_h$   &20.768(24)      & ~~1.2\\
$R_b$ &0.21642(73)       & ~~0.85\\
$R_c$&  0.1674(38)&    $-1.3$ \\
$A^l_{FB}$&  0.01701(95) & $~~0.8$ \\
$A_\tau$ &      0.1425(44)       & $-1.2$ \\
$A_e$   &0.1483(51) & $~~0.1$\\
$A^b_{FB}$ &    0.0988(20)  & $-2.3$ \\
$A^c_{FB}$&     0.0692(37)      & $-1.3$\\
$A_b$ (SLD direct)   & 0.911(25) & $-1.0$\\ 
$A_c$ (SLD direct)  &  0.630(26) & $-1.5$\\ 
$\sin^2\theta_{eff}({\rm\hbox{LEP-combined}})$ & 0.23192(23) &$ ~~2.1$\\
$A_{LR}\rightarrow  \sin^2\theta_{eff}$& 0.23099(26) & $-1.9$ \\
$m_W$ (GeV) (LEP2+p$\bar p$) & 80.419(38)    & $~~0.1$\\
$1-\frac{m^2_W}{m^2_Z}$ ($\nu$N) &  0.2255(21)  & $~~1.2$\\ 
$Q_W$ (Atomic PV in Cs) &  -72.06(44) & $~~2.5$\\
$m_t$ (GeV)     &174.3(5.1) & $~~0.1$\\
\hline
\end{tabular}
\end{center}
\end{table}

	For the analysis of electroweak data in the SM one starts from the input parameters: some of them,
$\alpha$, $G_F$ and $m_Z$, are very well measured, some other ones, $m_{f_{light}}$, $m_t$ and
$\alpha_s(m_Z)$  are only approximately determined while $m_H$ is largely unknown. With respect to
$m_t$ the situation has much improved since the CDF/D0 direct measurement of the top quark mass. From the input parameters
one computes the radiative corrections to a sufficient precision to match the experimental capabilities. Then
one compares the theoretical predictions and the data for the numerous observables which have been measured, checks the
consistency of the theory and derives constraints on $m_t$, $\alpha_s(m_Z)$ and hopefully also on $m_H$. 

	Some comments on the least known of the input parameters are now in order.The only practically
relevant terms where precise values of the light quark masses, $m_{f_{light}}$, are needed are those
related to the hadronic contribution to the photon vacuum polarization diagrams, that detemine
$\alpha(m_Z)$. This correction is of order 6$\%$, much larger than the accuracy of a few per mille of
the precision tests. Fortunately, one can use the actual data to in principle solve the related
ambiguity. But we shall see that the left over uncertainty is still one of the main sources of
theoretical error.
As is well known  \cite{radcorr}, the QED running coupling is given by:
\begin{equation}
\alpha(s) = \frac{\alpha}{1-\Delta \alpha(s)}
\label{1a}
\end{equation}
\begin{equation}	
\Delta \alpha(s) = \Pi(s) = \Pi_\gamma(0) - {\rm Re} \Pi_\gamma(s)
\label{2a}
\end{equation}
where $\Pi(s)$ is proportional to the sum of all 1-particle irreducible vacuum
polarization diagrams. In perturbation theory $\Delta\alpha(s)$ is given by:
\begin{equation}
\Delta \alpha(s) = \frac{\alpha}{3\pi} \sum_f Q^2_f N_{Cf}\left( \log
\frac{2}{m^2_f} - \frac{5}{3} \right)
\label{3a}
\end{equation}
where $N_{Cf} = 3$ for quarks and 1 for leptons. However, the perturbative formula
is only reliable for leptons, not for quarks (because of the unknown values of the
effective quark masses). Separating the leptonic, the light quark and the top
quark contributions to $\Delta\alpha(s)$ we have:
\begin{equation}
\Delta\alpha(s) = \Delta\alpha(s)_1 + \Delta\alpha(s)_h + \Delta\alpha(s)_t
\label{4a}
\end{equation}		
with:
\begin{equation}
\Delta\alpha(s)_1 = 0.0331421~;~~\Delta\alpha(s)_t =
\frac{\alpha}{3\pi}~\frac{4}{15}~\frac{m^2_Z}{m^2_t} = -0.000061
\label{5a}
\end{equation}
Note that in QED there is decoupling so that the top quark contribution approaches
zero in the large $m_t$ limit. For $\Delta\alpha(s)_h$ one can use eq.(\ref{2a}) and
the Cauchy theorem to obtain the representation:
\begin{equation}
\Delta\alpha(m^2_Z)_h = -\frac{\alpha m^2_Z}{3\pi}{\rm Re}
\int^\infty_{4m^2_\pi}\frac{ds}{s}~\frac{R(s)}{s-m^2_Z-i\epsilon}
\label{6a}
\end{equation}
where $R(s)$ is the familiar ratio of the hadronic to the pointlike $\ell^+\ell^-$
cross-section from photon exchange in $e^+e^-$ annihilation. At $s$ large one can
use the perturbative expansion for $R(s)$ while at small $s$ one can use the actual
data.  In recent years there has been a lot of activity on this subject \cite{ew}. 
A conservative value, directly obtained from the data, is given by
\beq
\alpha(m_Z)^{-1}=128.90\pm0.09~~~~~[\Delta\alpha(m^2_Z)_h=0.02804\pm0.00064]\label{8aa} \eeq  
As I said, for the derivation of this result the QCD theoretical prediction is actually used for large values of s where the
data do not exist. But the sensitivity of the dispersive integral to this region is strongly suppressed, so that no
important model dependence is introduced. More recently some analyses have appeared where one studied by how much the error
on $\alpha_s(m_Z)$ is reduced by using the QCD prediction down to $\sqrt{s}=m_\tau$, with the possible exception of the
regions around the charm and beauty thresholds. These attempts were motivated by the apparent success of QCD
predictions in $\tau$ decays, despite the low $\tau$ mass (note however that the relevant currents are V-A in $\tau$ decay
but V in the present case). One finds that the central value is not much changed while the error in eq.(\ref{8aa}) is
reduced but, of course, at the price of more model dependence. For example, one quoted value is:
\beq
\alpha(m_Z)^{-1}=128.913\pm0.035~~~~~[\Delta\alpha(m^2_Z)_h=0.027782\pm0.000254]\label{8ab} \eeq 
The data from BES and Daphne are expected to somewhat improve the accuracy.

As for the strong coupling $\alpha_s(m_Z)$ the world average central value is by now quite stable. The
error is going down because the dispersion among the different measurements is much smaller in the most
recent set of data. 
The error on the final average is taken by all authors between
$\pm$0.003 and
$\pm$0.005 depending on how conservative one is. In the following our reference value
will be \beq 
\alpha_s(m_Z) = 0.119\pm0.003 \label{9aa} \eeq

Finally a few words on the current status of the direct measurement of $m_t$. The present combined CDF/D0
result is
\beq 
m_t = 174.3\pm 5.1~GeV \label{10aa} 
\eeq
The error is so small by now that one is approaching a level
where a more careful investigation of the effects of colour rearrangement on the determination of $m_t$ will be
needed. One wants to determine the top quark mass, defined as the invariant mass of its decay products (i.e.
b+W+ gluons +
$\gamma$'s). However, due to the need of colour rearrangement, the top quark and its decay products cannot be
really isolated from the rest of the event. Some smearing of the mass distribution is induced by this colour
crosstalk which involves the decay products of the top, those of the antitop and also the fragments of the
incoming (anti)protons. A reliable quantitative computation of the smearing effect on the $m_t$ 
determination is difficult because of the importance of non perturbative effects. An induced error of
the order of 1 GeV on $m_t$ could reasonably be expected. So this problem is still not urgent. 

In order to appreciate the relative importance of the different sources of theoretical error for
precision tests of the SM, we report in table 2  a comparison for the most relevant observables.	What is important to stress
is that the ambiguity from $m_t$, once by far the largest one, is by now smaller than the error from $m_H$. We also see from
table 2 that the error from
$\Delta\alpha(m_Z)$ is expecially important for $\sin^2\theta_{eff}$  and, to a lesser extent, is also sizeable for
$\Gamma_Z$ and $\epsilon_3$.  
\begin{table}
 \caption{Errors from different sources: $\Delta^{exp}_{now}$    is
the present experimental error;
$\Delta\alpha^{-1}$ is the impact of $\Delta\alpha^{-1}=\pm0.09$;  $\Delta_{th}$
is the estimated theoretical error from higher orders; $\Delta m_t$ is from
$\Delta m_t =\pm 6 $GeV;
$\Delta m_H$ is from $\Delta m_H$ = 60--1000 GeV; $\Delta \alpha_s$ corresponds to
$\Delta \alpha_s=\pm0.003$. The epsilon parameters are defined in
sect.7.2.}
\label{tab2}
\begin{center}
\begin{tabular}{|l|l|l|l|l|l|l|}
\hline Parameter& $\Delta^{exp}_{now}$ & $\Delta \alpha^{-1}$ & $\Delta_{th}$ &
$\Delta m_t$ & $\Delta m_H$ & $\Delta \alpha_s$ \\
\hline
$\Gamma_Z$ (MeV) & $\pm$2.4 & $\pm$0.7 & $\pm$0.8 & $\pm$1.4 & $\pm$4.6 &
$\pm$1.7 \\
$\sigma_h$ (pb) & 37 & 1 & 4.3 & 3.3 & 4 & 17\\
$R_h \cdot 10^3$ & 24 & 4.3 & 5 & 2 & 13.5 & 20 \\
$\Gamma_l$ (keV) & 89 & 11 & 15 & 55 & 120 & 3.5\\
$A^l_{FB}\cdot 10^4$ & 9.5 & 4.2 & 1.3 & 3.3 & 13 & 0.18 \\
$\sin^2\theta\cdot 10^4$ & 1.7 & 2.3 & 0.8 & 1.9 & 7.5 & 0.1\\
$m_W$~(MeV) & 38 & 12 & 9 & 37 & 100& 2.2 \\
$R_b \cdot 10^4$ & 7.3 & 0.1 & 1 & 2.1 & 0.25 & 0\\
$\epsilon_1\cdot 10^3$ & 1.1 & & $\sim$0.1 & & & 0.2\\
$\epsilon_3\cdot 10^3$ & 1.0 & 0.5 & $\sim$0.1 & & & 0.12\\
$\epsilon_b\cdot 10^3$ & 1.8 & & $\sim$0.1 & & & 1\\
\hline
\end{tabular}
\end{center}
\end{table}

An important recent advance in the theory of radiative corrections is the calculation of the 
$o(g^4m^2_t/m^2_W)$ terms in $\sin^2\theta_{eff}$, $m_W$ and, more recently in $\delta\rho$ \cite{ew}. The result implies
a small but visible correction to the predicted values but expecially a seizable decrease of the ambiguity
from scheme dependence (a typical effect of truncation). These calculations are now implemented in the fitting codes used
in the analysis of LEP data. The fitted value of the Higgs mass is lowered by about
$30~GeV$ due to this effect. 

We now discuss fitting the data in the SM. As the mass of the top quark is now rather precisely known from CDF and D0 one
must distinguish two different types of fits. In one type one wants to answer the question: is $m_t$ from radiative
corrections in agreement with the direct measurement at the Tevatron? Similarly how does $m_W$ inferred from radiative
corrections compare with the direct measurements at the Tevatron and LEP2?  For answering these interesting but somewhat
limited questions, one must clearly exclude the measurements of
$m_t$ and $m_W$ from the input set of data. Fitting all other data in terms of
$m_t$,
$m_H$ and
$\alpha_s(m_Z)$ one finds the results shown in the second column of table 3~\cite{ew}. The extracted value of $m_t$ is
in good agreement with the direct measurement.In fact, as shown in the table 3, from all the electroweak data except the
direct production results on $m_t$ and $m_W$,  one finds
$m_t= 167\pm ^{11}_8 GeV$. There is a strong correlation between $m_t$ and $m_H$.
 In a more general type of fit, e.g. for determining the overall consistency
of the SM or to evaluate the best present estimate for some quantity, say $m_W$, one should of course not ignore the existing
direct determinations of $m_t$ and $m_W$. Then, from all the available data,  by fitting
$m_t$, $m_H$ and $\alpha_s(m_Z)$ one finds the values shown in the last column of table 3.
\begin{table}
\caption{ Standard Model fits of electroweak data.\label{tab3}}
\begin{center}
\begin{tabular}{|l|l|l|l|}
\hline Parameter & LEP(incl.$m_W$) &All but $m_W$, $m_t$  & All Data\\
\hline
$m_t$ (GeV) & 172$+14-11$ & 167$+11-8$ & $173.2\pm4.5$\\
$m_H$ (GeV) & 134$+268-81$ & 55$+84-27$ &77$+69-39$\\
$log[m_H(GeV)]$ & 2.13$+0.48-0.40$ &  1.74$+0.40-0.30$ &  1.88$+0.28-0.30$ \\
$\alpha_s(m_Z)$ & $0.120\pm0.003$ & $0.118\pm0.003$ & $0.118\pm0.003$ \\
$\chi^2/dof$ & 11/9 & 21/12 &23/15\\
\hline
\end{tabular}
\end{center}
\end{table}
 
This is the fit also referred to in table 1. The corresponding fitted values of
$\sin^2\theta_{eff}$ and $m_W$ are: 
\beq
\sin^2\theta_{eff} =0.23150\pm0.00016\nonumber;~~~~
                        m_W = 80.385\pm0.022 ~GeV \label{10car} 
\eeq 
The fitted value of $\sin^2\theta_{eff}$ is
practically identical to the LEP+SLD average. The error of 22 MeV on
$m_W$  clearly sets up a goal for the direct measurement of $m_W$ at LEP2, the Tevatron and the LHC.

The main lesson of the precision tests of the standard electroweak theory can be summarised as follows. It has
been checked that the couplings of quark and leptons to the weak gauge bosons $W^{\pm}$ and $Z$ are indeed
precisely those prescribed by the gauge symmetry. The accuracy of a few $0.1\%$ for these tests implies that, not
only the tree level, but also the structure of quantum corrections has been verified. To a lesser accuracy the
triple gauge vertices
$\gamma W^+ W^-$ and
$Z W^+ W^-$ have also been found in agreement with the specific prediction, at the tree level, of the $SU(2)\bigotimes U(1)$
gauge theory. This means that it has been verified that the gauge symmetry is indeed unbroken in the
vertices of the theory: the currents are indeed conserved. Yet there is obvious evidence that the symmetry is
otherwise badly broken in the masses. In fact the $SU(2)\bigotimes U(1)$ gauge symmetry forbids masses for all
the particles that have been sofar observed: quarks, leptons and gauge bosons. But of all these particles
only the photon and the gluons are massless ( protected by the $SU(3) \bigotimes U(1)_Q$ unbroken colour-electric charge gauge
symmetry), all other are massive (probably also the neutrinos). Thus the currents are conserved but the spectrum of particle
states is not symmetric. This is the definition of spontaneous symmetry breaking. The practical implementation of spontaneous
symmetry breaking in a gauge theory is via the Higgs mechanism. In the minimal SM one single
fundamental scalar Higgs isospin doublet is introduced and its vacuum expectation value v breaks the symmetry. All masses
are proportional to v, although the Yukawa couplings that multiply v in the
expression for the masses of quarks and leptons are distributed over a wide range. The Higgs sector is still very much
untested. The Higgs particle has not been found but its mass can well be heavier than the present direct lower limit
$m_H\gappeq113~GeV$  from LEP2 \footnote{this combined limit was presented by the LEP collaborations at the 2000
summer conferences}.  One knew from the beginning that the Higgs search would be difficult: being coupled in proportion to
masses one has first to produce heavy particles and then try to detect the Higgs (itself heavy) in their couplings. What has
been tested is the relation
$m^2_W=m^2_Z
\cos^2{\theta_W}$, modified by computable radiative corrections. This relation means that the effective Higgs (be it
fundamental or composite) is indeed a weak isospin doublet. 

We have seen that quantum corrections depend only
logaritmically on $m_H$.  In spite of this small sensitivity, the data are precise enough that one obtains a quantitative
indication of the mass range:
\cite{ew} $\log_{10}{m_H(GeV)}=1.88^{+0.28}_{-0.30}$ (or
$m_H=77^{+69}_{-39}$). This result on the Higgs mass is particularly remarkable. The value of
$\log_{10}{m_H(GeV)}$ is right on top of the small window between $\sim 2$ and $\sim 3$ which is allowed by the
direct limit, on the one side, and the theoretical upper limit on the Higgs mass in the minimal SM (see later),
$m_H\lappeq 600-800~GeV$, on the other side. If one had found a central value like $\gappeq 4$ the model would have
been discarded. Thus the whole picture of a perturbative theory with a fundamental Higgs is well supported by the
data on radiative corrections. It is important that there is a clear indication for a particularly light Higgs. This is
quite encouraging for the ongoing search for the Higgs particle. More in general, if the Higgs couplings are removed
from the lagrangian the resulting theory is non renormalisable. A cutoff $\Lambda$ must be introduced. In the quantum
corrections 
$\log{m_H}$ is then replaced by $\log{\Lambda}$ plus a constant. The precise determination of the associated finite
terms would be lost (that is, the value of the mass in the denominator in the argument of the logarithm). Thus the fact
that, from experiment, one finds $\log{m_H}\sim 2$ is a strong argument in favour of the specific form of the Higgs
mechanism as in the SM. A heavy Higgs would need some unfortunate conspiracy: the finite terms should accidentally
compensate for the heavy Higgs in the few key parameters of the radiative corrections (mainly $\epsilon_1$ and $\epsilon_3$).
Or additional new physics, for example in the form of effective contact terms added to the minimal SM lagrangian, should
accidentally do the compensation, which again needs some sort of conspiracy. 

\subsection{A More General Analysis of Electroweak Data}

We now discuss an update of the epsilon analysis \cite{ABC98} which is a method to look at the data in
a more general context than the SM. This is important to put constraints on extensions of the SM. The starting point is to
isolate from the data that part which is due to the purely weak radiative corrections. In fact 
the epsilon variables are defined in such a way that they are zero
in the
approximation when only effects from the SM at the tree level plus pure QED and pure QCD corrections are taken into account.
 This
very simple version of improved Born approximation is a good first approximation  according to the data and is independent of
$m_t$ and $m_H$. In fact the whole $m_t$ and $m_H$ dependence arises from weak loop corrections and therefore is only
contained in the epsilon variables. Thus the epsilons are extracted from the data without need of specifying
$m_t$ and $m_H$. But their predicted value in the SM or in any extension of it depend on $m_t$ and $m_H$.
This is to be compared with the competitor method based on the S, T, U variables. The latter
cannot be obtained from the data without specifying
$m_t$ and $m_H$ because they are defined as deviations from the complete SM prediction for specified $m_t$ and
$m_H$. Of course there are very many variables that vanish if pure weak loop corrections are neglected, at
least one for each relevant observable. Thus for a useful definition we choose a set of
representative observables that are used to parametrize those hot spots of the radiative corrections where
new physics effects are most likely to show up. These sensitive weak correction terms include vacuum
polarization diagrams which being potentially quadratically divergent are likely to contain possible non
decoupling effects (like the quadratic top quark mass dependence in the SM). There are three independent
vacuum polarization contributions. In the same spirit, one must add the $Z\rightarrow b \bar b$ vertex which also
includes a large top mass dependence. Thus altogether we consider four defining observables: one asymmetry, for
example
$A_{FB}^l$, (as representative of the set of measurements that lead to the determination of
$\sin^2\theta_{eff}$), one width (the leptonic width
$\Gamma_l$ is particularly suitable because it is practically independent of $\alpha_s$), $m_W$ and $R_b$.
Here lepton universality has been taken for granted, because the data show that it is
verified within the present accuracy. The
four variables,
$\epsilon_1$, $\epsilon_2$, $\epsilon_3$ and $\epsilon_b$ are defined in
correspondence with the set of observables  $A^{FB}_l$, $\Gamma_l$,
$m_W$, and $R_b$. The definition is so chosen that the quadratic top mass dependence is only
present  in
$\epsilon_1$ and
$\epsilon_b$, while the
$m_t$ dependence of
$\epsilon_2$ and $\epsilon_3$ is logarithmic. The definition of $\epsilon_1$ and $\epsilon_3$ is specified
in terms of $A^{FB}_l$ and $\Gamma_l$ only. Then adding $m_W$ or $R_b$ one obtains $\epsilon_2$ or
$\epsilon_b$. We now specify the relevant definitions in detail.

We start from the basic observables $m_W/m_Z$, $\Gamma_l$ and  $A^{FB}_l$ and $\Gamma_b$. From these four
quantities one can isolate the corresponding dynamically significant corrections $\Delta r_W$, $\Delta \rho$, 
$\Delta k$ and $\epsilon_b$, which  contain the small effects one is trying to disentangle and are defined in
the following. First we introduce $\Delta r_W$ as obtained from $m_W/m_Z$ by the relation:
\beq
(1-\frac{m_W^2}{m_Z^2}) \frac{m_W^2}{m_Z^2}~=~\frac{\pi \alpha(m_Z)}{\sqrt{2} G_F m_Z^2 (1-\Delta r_W)}
\label{1n}
\eeq
Here $\alpha(m_Z)~=~\alpha /(1-\Delta \alpha)$ is fixed to the central value 1/128.90 so that the effect of
the running of $\alpha$ due to known physics is extracted from $1-\Delta r = (1- \Delta \alpha)(1- \Delta
r_W)$. In fact, the error on $1/\alpha(m_Z)$, as given in eq.(\ref{8aa}) would then affect $\Delta r_W$.
In order to define $\Delta
\rho$ and 
$\Delta k$ we
consider the effective vector and axial-vector couplings $g_V$ and $g_A$ of the on-shell Z to charged leptons,
given by the formulae:
\bea
\Gamma_l~&=&~\frac{G_F m^3_Z}{6\pi \sqrt{2}}(g^2_V+g_A^2) (1+\frac{3 \alpha}{4 \pi}), \nonumber \\
A_l^{FB}(\sqrt{s}&=&m_Z)~=~\frac{3g^2_Vg^2_A}{(g^2_V+g_A^2)^2}~=~\frac{3x^2}{(1+x^2)^2}. \label{2nn}
\eea
Note that $\Gamma_l$ stands for the inclusive partial width $\Gamma(Z\rightarrow l\bar l + \rm{photons})$. We
stress the following points. First, we have extracted from $(g^2_V+g_A^2)$ the factor $(1 + 3\alpha /4 \pi )$
which is induced in $\Gamma_l$ from final state radiation. Second, by the  asymmetry at the peak in
eq.(\ref{2nn}) we mean the quantity which is commonly referred to by the LEP experiments (denoted as $A^0_{FB}$
in ref.\cite{ew}), which is corrected for all QED effects, including initial and final state radiation and
also for the effect of the imaginary part of the $\gamma$ vacuum polarization  diagram. In terms of $g_A$ and
$x= g_V /g_A$, the quantities $\Delta \rho$ and 
$\Delta k$ 
are given by:
\bea
g_A~=~-\frac{\sqrt{\rho}}{2}~\sim~-\frac{1}{2}(1+\frac{\Delta \rho}{2}), \nonumber \\
x~=~\frac{g_V}{g_A}~=~1-4\sin^2\theta_{eff}~=~1-4(1+\Delta k) s_0^2.\label{3n}
\eea
Here $s_0^2$ is $\sin^2\theta_{eff}$
before non pure-QED corrections, given by:
\beq
s_0^2 c_0^2~=~ \frac{\pi \alpha(m_Z)}{\sqrt{2} G_F m_Z^2} \label{4n}
\eeq
with  $c_0^2~=~1-s_0^2$ ($s_0^2 = 0.231095$ for $m_Z~=~91.188~GeV$).
	
We now define $\epsilon_b$ from $\Gamma_b$, the inclusive partial width for $Z\rightarrow b \bar b$ according to
the relation
\beq
\Gamma_b~=~\frac{G_F m^3_Z}{6\pi \sqrt{2}}\beta (\frac{3-\beta^2}{2} g^2_{bV}~+~\beta^2 g^2_{bA}) N_C R_{QCD}
(1+\frac{\alpha}{12\pi}) \label{5n}
\eeq
where $N_C=3$ is the number of colours, $\beta=\sqrt{1-4m_b^2/m^2_Z}$, with $m_b=4.7~$
GeV, $R_{QCD}$ is the QCD correction factor given by
\beq
R_{QCD}~=~ 1~+~1.2a~-~1.1a^2~-~13a^3~;~~~a~=~\frac{\alpha_s(m_Z)}{\pi} \label{6n}
\eeq
and $g_{bV}$ and $g_{b A}$ are specified as follows
\bea
g_{bA}~=~-\frac{1}{2}(1+\frac{\Delta \rho}{2})(1+\epsilon_b), \nonumber\\
\frac{g_{bV}}{g_{bA}}~=~\frac{1-4/3\sin^2\theta_{eff}+\epsilon_b}{1+\epsilon_b}.\label{7n}
\eea
This is clearly not the most general deviation from the SM in the $Z\rightarrow b \bar b$ but $\epsilon_b$ is
closely related to the quantity  $-Re(\delta_{b-vertex})$ where the large
$m_t$ corrections are located in the SM.

As is well known, in the SM the quantities $\Delta r_W$, $\Delta \rho$, $\Delta k$  and $\epsilon_b$, for
sufficiently  large $m_t$, are all dominated by  quadratic terms in $m_t$ of order $G_Fm^2_t$.   As new physics
can  more easily be disentangled if not masked by large conventional $m_t$ effects, it is convenient to keep
$\Delta \rho$ and $\epsilon_b$ while trading $\Delta r_W$
and 
$\Delta k$ for two quantities with no contributions of order $G_Fm^2_t$. We thus introduce the
following linear combinations:
\bea
\epsilon_1 &=& \Delta \rho, \nonumber \\
\epsilon_2 &=& c^2_0 \Delta \rho~+~\frac{s^2_0 \Delta r_W}{c^2_0-s^2_0}~-~2s^2_0 \Delta k, \nonumber\\
\epsilon_3 &=& c^2_0 \Delta \rho~+~(c^2_0-s^2_0) \Delta k. \label{8n}
\eea
The quantities $\epsilon_2$ and $\epsilon_3$ no longer contain terms of order $G_Fm^2_t$
 but only logarithmic
terms in $m_t$. The leading terms for large Higgs mass, which are logarithmic, are contained in
$\epsilon_1$ and $\epsilon_3$. In the Standard Model one has the following "large"
asymptotic contributions:
\bea
\epsilon_1 &=& \frac{3G_F m_t^2}{8 \pi^2 \sqrt{2}}~-~\frac{3G_F m_W^2}{4 \pi^2 \sqrt{2}}
 \tan^2{\theta_W}
\ln\frac{m_H}{m_Z}~+....,\nonumber \\
\epsilon_2 &=& -\frac{G_F m_W^2}{2 \pi^2 \sqrt{2}}\ln\frac{m_t}{m_Z}~+....,\nonumber \\
\epsilon_3 &=& \frac{G_F m_W^2}{12 \pi^2 \sqrt{2}}\ln\frac{m_H}{m_Z}~-~\frac{G_F m_W^2}{6 \pi^2
\sqrt{2}}\ln\frac{m_t}{m_Z}....,\nonumber \\
\epsilon_b &=& -\frac{G_F m_t^2}{4 \pi^2 \sqrt{2}}~+.... \label{9n}
\eea

The relations between the basic observables and the epsilons can be linearised, leading to the
approximate formulae
\bea
\frac{m_W^2}{m_Z^2}~&=&~\frac{m_W^2}{m_Z^2}\vert_B (1+ 1.43\epsilon_1 - 1.00\epsilon_2 - 0.86\epsilon_3),
\nonumber \\
\Gamma_l~&=&~\Gamma_l\vert_B (1+ 1.20\epsilon_1 - 0.26\epsilon_3),
\nonumber \\
A_l^{FB}~&=&~A_l^{FB} \vert_B (1+ 34.72\epsilon_1 - 45.15\epsilon_3),
\nonumber \\
\Gamma_b~&=&~\Gamma_b\vert_B (1+ 1.42\epsilon_1 - 0.54\epsilon_3 + 2.29\epsilon_b).  \label{10n}
\eea
The  Born approximations, as defined above, depend on $\alpha_s(m_Z)$ and also on $\alpha(m_Z)$. Defining
\beq
\delta \alpha_s~=~\frac{\alpha_s(m_Z)-0.119}{\pi};~~~\delta
\alpha~=~\frac{\alpha(m_Z)-\frac{1}{128.90}}{\alpha},~~~~ \label{11n}
\eeq
we have
\bea
\frac{m_W^2}{m_Z^2}\vert_B~&=&~0.768905(1-0.40\delta \alpha), \nonumber \\
\Gamma_l\vert_B~&=&~83.563(1-0.19\delta \alpha) \rm{MeV}, \nonumber \\
A_l^{FB} \vert_B~&=&~0.01696(1-34\delta \alpha), \nonumber \\
\Gamma_b\vert_B~&=&~379.8(1+1.0\delta \alpha_s-0.42\delta \alpha). \label{12nn}
\eea 
Note that the dependence on $\delta \alpha_s$ for $\Gamma_b\vert_B$, shown in eq.(\ref{12nn}), is not simply the
one loop result for $m_b=0$ but a combined effective shift which takes into account both finite mass effects
and the contribution of the known higher order terms.

\begin{table}
\caption{Values of the epsilons in the SM as functions of $m_t$ and
$m_H$ as obtained from recent versions of ZFITTER  and TOPAZ0.
These values (in
$10^{-3}$ units) are obtained for
$\alpha_s(m_Z)$ = 0.119,
$\alpha(m_Z)$ = 1/128.90, but the theoretical predictions are essentially
independent of
$\alpha_s(m_Z)$ and $\alpha(m_Z)$}
\label{tab4}
\begin{center}
\begin{tabular}{|c|l|l|l|l|l|l|l|l|l|c|}
\hline
$m_t$ & \multicolumn{3}{|c|}{$\epsilon_1$}&\multicolumn{3}{|c|}{$\epsilon_2$}
&\multicolumn{3}{|c|}{$\epsilon_3$}&$\epsilon_b$\\ (GeV)& \multicolumn{3}{|c|}
{$m_H$ (GeV) =} &  \multicolumn{3}{|c|} {$m_H$ (GeV) =} & \multicolumn{3}{|c|}
{$m_H$ (GeV) =} & All {$m_H$}\\ & 70 & 300 & 1000 & 70 & 300 & 1000 & 70 & 300 &
1000 &\\
\hline 150      &3.55&  2.86    & 1.72 &        $-$6.85 &       $-$6.46 &       $-$5.95 &        4.98    & 6.22 &        6.81 &
$-$4.50 \\ 160 &        4.37 &  3.66 &  2.50 &  $-$7.12 &       $-$6.72 &        $-$6.20 &       4.96 &   6.18 &
6.75 &  $-$5.31
\\
 170 &  5.26 &  4.52 &  3.32 &  $-$7.43 &        $-$7.01 &        $-$6.49 &       4.94 &  6.14 &  6.69 &
$-$6.17\\
 180 &  6.19 &   5.42 &  4.18 &   $-$7.77 &       $-$7.35 &       $-$6.82 &       4.91 &  6.09 &  6.61 &
$-$7.08\\
 190 &  7.18 &   6.35 &  5.09 &  $-$8.15 &       $-$7.75 &       $-$7.20 &       4.89 &  6.03 &   6.52 &
$-$8.03\\
 200 &  8.22 &  7.34 &  6.04 &   $-$8.59 &       $-$8.18 &       $-$7.63 &       4.87 &  5.97 &  6.43 &
$-$9.01\\
\hline
\end{tabular}
\end{center}
\end{table}

The important property of the epsilons is that, in the Standard Model, for all observables at the Z pole, the
whole dependence on $m_t$ (and $m_H$) arising from one-loop diagrams only enters through the epsilons. The same
is actually true, at the relevant level of precision, for all higher order $m_t$-dependent corrections.
Actually, the only residual $m_t$ dependence of the various observables not included in the epsilons is in the
terms of order $\alpha_s^2(m_Z)$ in the pure QCD correction factors to the hadronic widths. But this
one is quantitatively irrelevant, especially in view of the errors connected to the uncertainty on the value of
$\alpha_s(m_Z)$. The theoretical values of the epsilons in the SM from state of the art radiative corrections are given in
table 4. It is important to remark that the theoretical values of the epsilons in the SM, as given in table 4, are not
affected, at the percent level or so, by reasonable variations of
$\alpha_s(m_Z)$ and/or
$\alpha(m_Z)$ around their central values. By our definitions, in fact,  no terms of order $\alpha_s^n(m_Z)$ or $\alpha \ln{m_Z/m}$
contribute to the epsilons.  In terms of the epsilons, the following expressions hold, within the SM, for the
various precision observables
\bea
\Gamma_T~&=&~\Gamma_{T0}(1+1.35\epsilon_1-0.46\epsilon_3+0.35\epsilon_b), \nonumber\\
R~&=&~R_0(1+0.28\epsilon_1-0.36\epsilon_3+0.50\epsilon_b), \nonumber\\
\sigma_h~&=&~\sigma_{h0}(1-0.03\epsilon_1+0.04\epsilon_3-0.20\epsilon_b), \nonumber\\
x~&=&~x_0(1+17.6\epsilon_1-22.9\epsilon_3), \nonumber\\
R_b~&=&~R_{b0}(1-0.06\epsilon_1+0.07\epsilon_3+1.79\epsilon_b). \label{13n}
\eea
where x=$g_V/g_A$ as obtained from $A_l^{FB}$ . The quantities in eqs.(\ref{10n}),(\ref{13n}) are clearly not
independent and the redundant information is reported for convenience. By comparison with the computed radiative corrections we
obtain
\bea
\Gamma_{T0}~&=&~2489.46(1+0.73\delta \alpha_s-0.35\delta \alpha)~MeV,\nonumber \\
R_0~&=&~20.8228(1+1.05\delta \alpha_s-0.28\delta \alpha),\nonumber \\
\sigma_{h0}~&=&~41.420(1-0.41\delta \alpha_s+0.03\delta \alpha)~nb,\nonumber \\
x_0~&=&~0.075619-1.32\delta \alpha,\nonumber \\
R_{b0}~&=&~0.2182355.\label{14nn}
\eea
Note that  the quantities in eqs.(\ref{14nn}) should not be confused, at least in principle, with the
corresponding Born approximations, due to small "non universal" electroweak corrections. In practice, at the
relevant level of approximation, the difference between the two corresponding quantities is in any case
significantly smaller than the present experimental error.

In principle, any four observables could have been picked up as defining variables. 
In practice we choose those that have a more clear physical significance and are more effective in the
determination of the epsilons. In fact,  since $\Gamma_b$ is actually measured by $R_b$ (which is nearly
insensitive to $\alpha_s$), it is preferable to use directly $R_b$  itself as defining variable, as we shall do
hereafter. In practice, since the value in eq.(\ref{14nn}) is practically indistinguishable from the Born
approximation of $R_b$, this determines no change in any of the equations given above but simply requires the
corresponding replacement among the defining relations of the epsilons.

The values of the epsilons as obtained from the defining variables $m_W$, $\Gamma_l$, $A^{FB}_l$ and $R_b$
are shown in the first column of table 5.
\begin{center}
\begin{table}
 \caption{Experimental values of the epsilons in the SM from different sets of data.
These values (in
$10^{-3}$ units) are obtained for
$\alpha_s(m_Z) = 0.119\pm0.003$,
$\alpha(m_Z)^{-1} = 128.913\pm0.035$, the corresponding uncertainties being
 included in the quoted errors}
 \label{tab5}
\vglue.3cm
\begin{tabular}{|l|l|l|l|l|}
\hline $\epsilon~~~10^3$  &Only def. quantities &All asymmetries &All High Energy & All Data\\
\hline
$\epsilon_1~10^3$ &$4.1\pm1.2$ &$4.3\pm1.2$ &$3.9\pm1.1$  &$3.2\pm1.1$ \\
$\epsilon_2~10^3$ &$-8.35\pm1.6$ &$-9.0\pm1.4$ &$-9.3\pm1.5$  &$-9.7\pm1.5$ \\
$\epsilon_3~10^3$ &$3.4\pm1.8$ &$4.5\pm1.1$ &$4.2\pm1.0$  &$3.5\pm1.0$ \\
$\epsilon_b~10^3$ &$-3.7\pm1.9$ &$-3.8\pm1.9$ &$-4.4\pm1.8$  &$-4.0\pm1.8$  \\
\hline
\end{tabular}
\end{table}
\end{center}
\vglue.3cm
To proceed further and include other measured observables in the analysis we need to make some
dynamical assumptions. The minimum amount  of model dependence is introduced by including other purely
leptonic quantities at the Z pole such as $A_{\tau}$, $A_e$ (measured  from the angular
dependence of the $\tau$ polarization) and $A_{LR}$ (measured by SLD). For this step, one is simply
assuming that the different leptonic asymmetries are equivalent measurements of $\sin^2\theta_{eff}$. We add, as usual, 
the measure of
$A^{FB}_b$ because this observable is dominantly sensitive to the leptonic vertex. We then use the combined value
of $\sin^2\theta_{eff}$ obtained from the whole set of asymmetries measured at LEP and SLC given in eq.(\ref{8}). At this stage the
best values of the epsilons are shown in the second column of table 5. In figs. 4-7  we report the 1$\sigma$ ellipses in the
indicated
$\epsilon_i$-$\epsilon_j$ planes that correspond to this set of
input data.
                                                      
\begin{figure}
\hglue2.0cm
\epsfig{figure=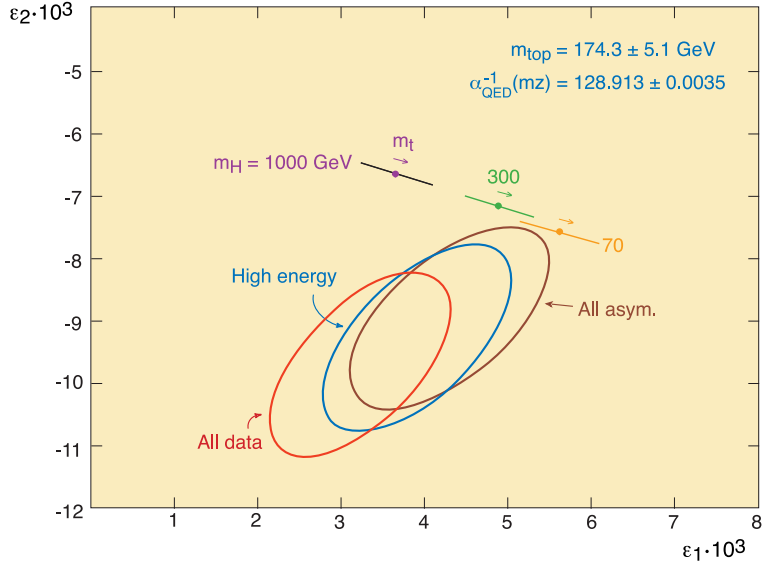,width=10cm}
\caption[ ]{Data vs theory in the $\epsilon_2$-$\epsilon_1$ plane. The origin point corresponds to the "Born"
approximation obtained from the SM at tree level plus pure QED and pure QCD corrections. The predictions of the
full SM  are shown for $m_H$ = 70, 300 and 1000 GeV and
$m_t=174.3\pm5.5~GeV$ (a segment for each $m_H$ with the arrow showing the direction of 
$m_t$ increasing from
$-1\sigma$ to $+1\sigma$). The three
$1-\sigma$ ellipses ($38\%$ probability contours) are obtained from a) "All Asymm." :$\Gamma_l$, $m_W$ and
$\sin^2\theta_{eff}$ as obtained from the combined asymmetries (the value in
eq. (\ref{ew8})); b) "All High
En.": the same as in a) plus all the hadronic variables at the Z; c) "All Data": the same as in b) plus the low
energy data}
\end{figure}

\begin{figure}
\hglue2.0cm
\epsfig{figure=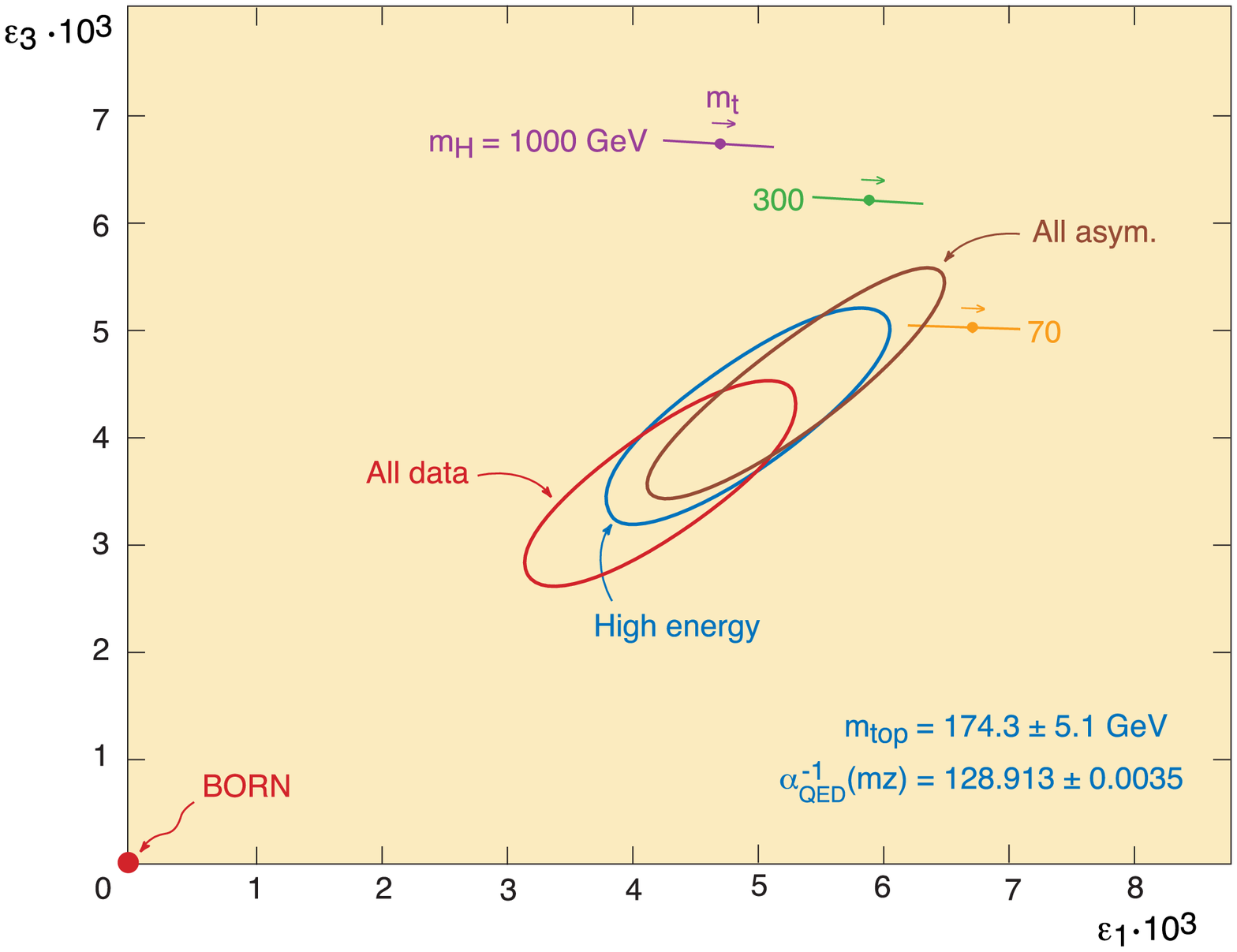,width=10cm}
\caption[ ]{Data vs theory in the $\epsilon_3 - \epsilon_1$ plane (notations as in fig. 4)}
\end{figure}

\begin{figure}
\hglue2.0cm
\epsfig{figure=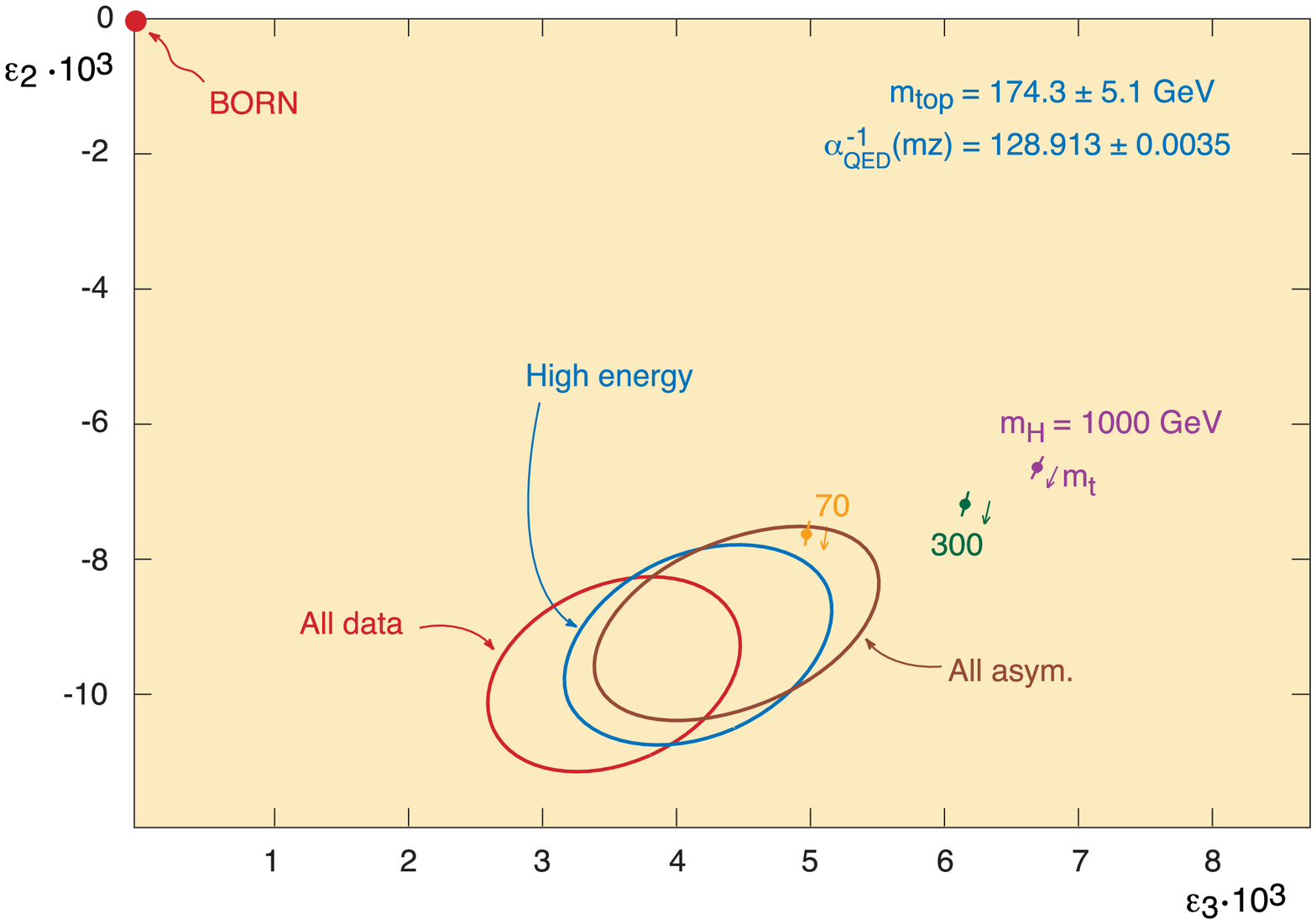,width=10cm}
\caption[ ]{Data vs Theory in the $\epsilon_2 - \epsilon_3$ plane (notations as in fig. 4)}
\end{figure}

\begin{figure}
\hglue2.0cm
\epsfig{figure=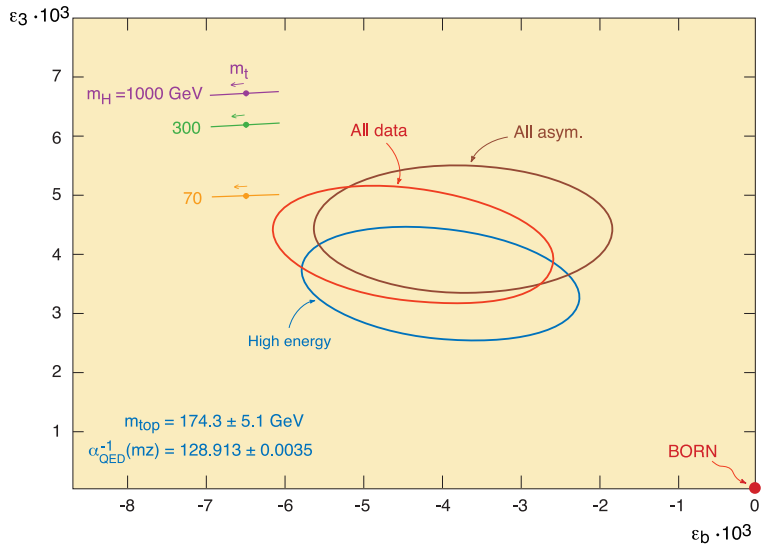,width=10cm}
\caption[ ]{Data vs theory in the $\epsilon_b - \epsilon_3$ plane (notations as in fig. 4)}
\end{figure}

	All observables measured on the Z peak at LEP can be included in the analysis provided that we assume
that all deviations from the SM are only contained in vacuum polarization diagrams (without demanding
a truncation of the $q^2$ dependence of the corresponding functions) and/or the $Z\rightarrow b\bar
b$  vertex. From a global fit of the data on $m_W$,  $\Gamma_T$,  $R_h$, $\sigma_h$,  $R_b$ and
$\sin^2\theta_{eff}$ (for LEP data, we have taken the correlation matrix for $\Gamma_T$,  $R_h$ and
$\sigma_h$ given by the LEP experiments \cite{ew}, while we have considered the additional information
on $R_b$ and $\sin^2\theta_{eff}$  as independent) we obtain the values shown in the third column of table
6. The comparison of theory and experiment at this stage is also shown in figs. 4-7. 

	 To include in our analysis lower energy observables as well, a stronger hypothesis needs to be
made: vacuum polarization diagrams are allowed to vary from the SM  only in their constant and first
derivative terms in a $q^2$ expansion. In such a case, one can, for example, add to the
analysis the ratio
$R_\nu$ of neutral to charged current processes in deep inelastic neutrino scattering on nuclei, the "weak charge" $Q_W$ 
measured in atomic parity violation experiments on Cs  and the measurement of $g_V/g_A$ from $\nu_\mu e$ scattering. In this
way one obtains  the global fit given in the fourth column of table 5 and shown in figs. 4-7. In fig. 8 we see the ellipse in
the $\epsilon _1$-$\epsilon _3$ plane that is obtained from the low energy data by themselves, in comparison with the results
from high energy data. We clearly see the effect of $2.5\sigma$ deviation from the SM fit of the measured parity violation in
atomic physics. It can be shown that the data on neutrino scattering fix the slope of the ellipse major axis, which is in
agreement with the high energy data. The atomic parity violation fix the center of the ellipse, which is instead displaced.
 
\begin{figure}
\hglue2.0cm
\epsfig{figure=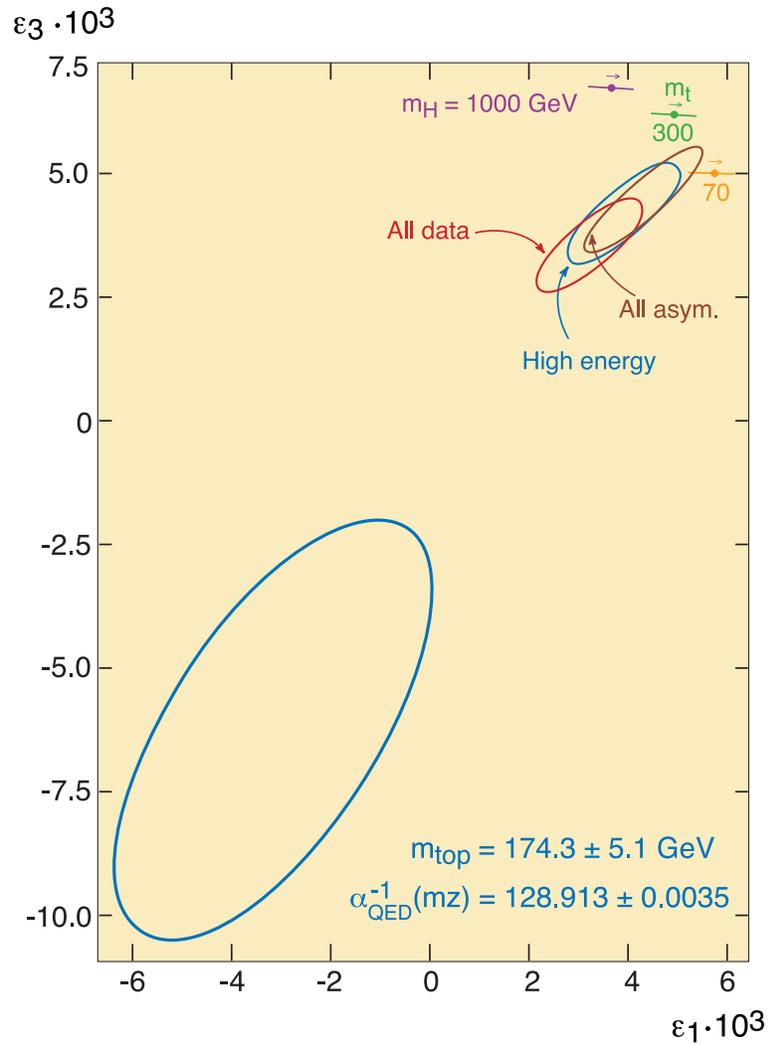,width=10cm}
\caption[ ]{Data vs theory in the $\epsilon_3 - \epsilon_1$ plane (notations as in fig. 4).  The ellipse
from the low energy data only is compared with that from high energy data and with the SM predictions.}
\end{figure}

The best
values of the
$\epsilon$'s from all the data are given in the last column of table 5.

Note that the ambiguity on the value of
$\delta\alpha^{-1}(m_Z) =\pm0.035$ (or $\pm 0.09$) corresponds to an uncertainty on
$\epsilon_3$ (the other epsilons are not much affected) given by $\Delta\epsilon_3~10^3 =\pm0.25$ (or $\pm 0.6$). Thus the
theoretical error is still confortably less than the experimental error. In fig.9 we present a summary of the
experimental values of the epsilons as compared to the SM predictions as functions of $m_t$ and $m_H$, which
shows agreement within $1\sigma$, but the central value of $\epsilon_1$, $\epsilon_2$ and $\epsilon_3$ are all a little bit 
low, while the central value of $\epsilon_b$ is shifted upward with respect to the SM as a
consequence of the still imperfect matching of $R_b$.

\begin{figure}
\hglue3.0cm
\epsfig{figure=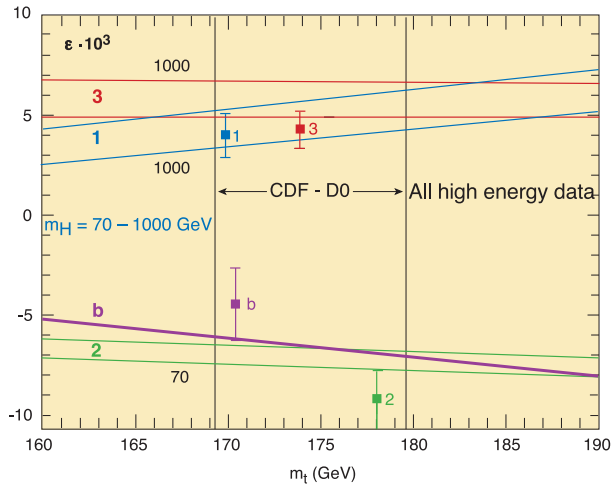,width=8cm}
\caption[ ]{The bands (labeled by the $\epsilon$ index) are the predicted values of the epsilons in the SM as functions of
$m_t$ for
$m_H~=~70-1000$ GeV (the $m_H$ value corresponding to one edge of the band is indicated). The CDF/D0 experimental
1-$\sigma$ range of $m_t$ is shown. The esperimental results for the epsilons from all data are displayed (from the last
column of table 5). The position of the data on the $m_t$ axis has been arbitrarily chosen and has no particular
meaning.}
\end{figure} 

A number of
interesting features are clearly visible from figs.5-11. First, the good agreement with the SM and the
evidence for weak corrections, measured by the distance of the data from the improved Born approximation point
(based on tree level SM plus pure QED or QCD corrections). There is by now a solid evidence for departures from
the improved Born approximation where all the epsilons vanish. In other words a clear evidence for the pure
weak radiative corrections has been obtained and one is sensitive to the various components of these
radiative corrections. For example, some authors  have studied the sensitivity of the data to a
particularly interesting subset of the weak radiative corrections, i.e. the purely bosonic part. These terms
arise from virtual exchange of gauge bosons and Higgses. The result is that indeed the measurements are
sufficiently precise to require the presence of these contributions in order to fit the data. Second, the
general results of the SM fits are reobtained from a different perspective. We see the preference for light
Higgs manifested by the tendency for
$\epsilon_3$ to be rather on the low side. Since $\epsilon_3$ is practically independent of $m_t$, its low value
demands $m_H$ small. If the Higgs is light then the preferred value of
$m_t$ is slightly lower than the Tevatron result (which in the epsilon analysis is not included among the input
data). This is because also the value of $\epsilon_1\equiv \delta \rho$, which is determined by the widths, in
particular by the leptonic width, is somewhat low. In fact
$\epsilon_1$ increases with $m_t$ and, at fixed $m_t$, decreases with $m_H$, so that for small $m_H$ the low
central value of $\epsilon_1$ pushes $m_t$ down. Note that also the central value of $\epsilon_2$ is on
the low side, because the experimental value of $m_W$ is a little bit too large. Finally, we see that adding the
hadronic quantities or the low energy observables hardly makes a difference in the
$\epsilon_i$-$\epsilon_j$ plots with respect to the case with only the leptonic variables being included (the
ellipse denoted by "All Asymm."). But, for example for the
$\epsilon_1$-$\epsilon_3$ plot, while the leptonic ellipse contains the same information as one could obtain from a
$\sin^2\theta_{eff}$ vs $\Gamma_l$ plot, the content of the other two ellipses is much larger because it
shows that the hadronic as well as the low energy quantities match the leptonic variables without need of any new
physics. Note that the experimental values of $\epsilon_1$ and
$\epsilon_3$ when the hadronic quantities are included also depend on the input value of $\alpha_s$ specified in table 5.

The good agreement of the fitted epsilon values with the SM impose strong constraints on possible forms of new physics.
Consider, for example, new quarks or leptons. Mass splitted multiplets contribute to $\Delta\epsilon_1$, in analogy to
the t-b quark doublet. Recall that $\Delta\epsilon_1\sim+9.5~10^{-3} $ for the t-b doublet, which is about
$10~\sigma$'s in terms of the present error. Even mass degenerate multiplets are strongly constrained. They
contribute to $\Delta\epsilon_3$ according to
\beq
\Delta\epsilon_3 \sim N_C \frac{G_Fm_W^2}{8\pi^2\sqrt{2}}\frac{4}{3}(T_{3L}-T_{3R})^2 \label{onehundredtwentyeight}\\
\eeq
 For example a new left-handed quark doublet, degenerate in mass, would contribute $\Delta\epsilon_3\sim +1.3~
10^{-3}$, that is more than one $\sigma$, but in the wrong direction, in the sense that the experimental value of
$\epsilon_3$ favours a displacement, if any, with negative sign. Only vector fermions $(T_{3L}=T_{3R})$ are not
constrained. In particular, naive technicolour models \cite{chi}, that introduce several new technifermions, are strongly
disfavoured because they tend to produce large corrections with the wrong sign to $\epsilon_1$,
$\epsilon_3$ and also to $\epsilon_b$.

\section{Why we do not Believe in the SM}
\subsection{Conceptual Problems}

Given the striking success of the SM why are we not satisfied with that theory? Why not just find the Higgs
particle, for completeness, and declare that particle physics is closed? The main reason is that there are
strong conceptual indications for physics beyond the SM. 

	It is considered highly unplausible that the origin of the electro-weak symmetry breaking can be explained by
the standard Higgs mechanism, without accompanying new phenomena. New physics should be manifest at energies in
the TeV domain. This conclusion follows fron an extrapolation of the SM at very high energies. The computed
behaviour of the $SU(3)\otimes SU(2)\otimes U(1)$ couplings with energy clearly points towards the
unification of the electro-weak and strong forces (Grand Unified Theories: GUT's) at scales of energy
$M_{GUT}\sim  10^{14}-10^{16}~ GeV$ which are close to the scale of quantum gravity, $M_{Pl}\sim 10^{19}~ GeV$
\cite{ross}.  One can also imagine  a unified theory of all interactions also including gravity (at
present superstrings provide the best attempt at such a theory). Thus GUT's and the realm of quantum gravity set a
very distant energy horizon that modern particle theory cannot anymore ignore. Can the SM without new physics be
valid up to such large energies? This appears unlikely because the structure of the SM could not naturally
explain the relative smallness of the weak scale of mass, set by the Higgs mechanism at $\mu\sim
1/\sqrt{G_F}\sim  250~ GeV$  with $G_F$ being the Fermi coupling constant. This so-called hierarchy problem is related to the
presence of fundamental scalar fields in the theory with quadratic mass divergences and no protective extra symmetry at
$\mu=0$. For fermion masses, first, the divergences are logaritmic and, second, they are forbidden by the $SU(2)\bigotimes
U(1)$ gauge symmetry plus the fact that at
$m=0$ an additional symmetry, i.e. chiral  symmetry, is restored. Here, when talking of divergences we are not
worried of actual infinities. The theory is renormalisable and finite once the dependence on the cut off is
absorbed in a redefinition of masses and couplings. Rather the hierarchy problem is one of naturalness. If we
consider the cut off as a manifestation of new physics that will modify the theory at large energy scales, then it
is relevant to look at the dependence of physical quantities on the cut off and to demand that no unexplained
enormously accurate cancellations arise. 

According to the above argument the observed value of $\mu\sim 250~ GeV$ is indicative of the existence of new
physics nearby. There are two main possibilities. Either there exist fundamental scalar Higgses but the theory
is stabilised by supersymmetry, the boson-fermion symmetry that would downgrade the bosonic degree of divergence from
quadratic to logarithmic. For approximate supersymmetry the cut off is replaced by the splitting between the
normal particles and their supersymmetric partners. Then naturalness demands that this splitting (times the
size of the weak gauge coupling) is of the order of the weak scale of mass, i.e. the separation within
supermultiplets should be of the order of no more than a few TeV. In this case the masses of most supersymmetric
partners of the known particles, a very large managerie of states, would fall, at least in part, in the discovery
reach of the LHC. There are consistent, fully formulated field theories constructed on the basis of this idea, the
simplest one being the MSSM \cite{43}. As already mentioned, all normal observed states are those whose masses are
forbidden in the limit of exact
$SU(2)\otimes U(1)$. Instead for all SUSY partners the masses are allowed in that limit. Thus when
supersymmetry is broken in the TeV range but $SU(2)\otimes U(1)$ is intact only s-partners take mass while all
normal particles remain massless. Only at the lower weak scale the masses of ordinary particles are generated.
Thus a simple criterium exists to understand the difference between particles and s-particles.

	The other main avenue is compositeness of some sort. The Higgs boson is not elementary but either a bound
state of fermions or a condensate, due to a new strong force, much stronger than the usual strong interactions,
responsible for the attraction. A plethora of new "hadrons", bound by the new strong force would  exist in the
LHC range. A serious problem for this idea is that nobody sofar has been  able to build up a realistic model
along these lines, but that could eventually be explained by a lack of ingenuity on the theorists side. The
most appealing examples are technicolor theories \cite{chi}. These models were inspired by the
breaking of chiral symmetry in massless QCD induced by quark condensates. In the case of the electroweak
breaking new heavy techniquarks must be introduced and the scale analogous to $\Lambda_{QCD}$ must be about
three orders of magnitude larger. The presence of such a large force relatively nearby has a strong tendency to
clash with the results of the electroweak precision tests. New versions have been developed to overcome the
negative response of the data, but models are far from offering a realistic picture.

Are there other ways to solve the hierarchy problem? Recently an exotic way was proposed \cite{hall}. The idea is
that perhaps the scale of gravity is only apparently so large. It has been shown that it is in principle possible to bring
down the scale of gravity in the multi TeV energy range. This can happen if one assumes the existence of extra space
dimensions with sufficiently large compactification radius, with the graviton propagating in all dimensions, while ordinary
gauge interactions are trapped on a four dimensional wall. The corresponding modification of gravity at submillimetric
distances is compatible with existing limits. The vicinity of the decompactification scale can manifest itself in high energy
processes at $e^+e^-$ and hadron colliders where gravitons can be produced and appear as missing energy. This very
speculative scenario is certainly interesting especially as a stimulus to look for specific signals. But does not appear
as particularly plausible because some large compactification scale have to be ad hoc introduced and large ratios of scales
still remain (e.g. the scale where gravity changes behaviour and the weak scale and largely different compactification
scales like the depth of the wall and the radius of the bulk). In addition all the positive hints we have in favour of the
ordinary picture of GUTs from coupling unification, neutrino masses, dark matter and so on would be emptied. Finally early time
cosmology should be rewritten.  

The hierarchy problem is certainly not the only conceptual problem of the SM. There are many more: the
proliferation of parameters, the mysterious pattern of fermion masses and so on. But while most of these
problems can be postponed to the final theory that will take over at very large energies, of order $M_{GUT}$ or
$M_{Pl}$, the hierarchy problem arises from the unstability of the low energy theory and requires a solution at
relatively low energies. 

A supersymmetric extension of the SM provides a way out which is well defined,
computable and that preserves all virtues of the SM.  The necessary SUSY breaking can be introduced through soft
terms that do not spoil the good convergence properties of the theory. Precisely those terms arise from
supergravity when it is spontaneoulsly broken in a hidden sector. This is the case in the Minimal
Supersymmetric Standard Model (MSSM) \cite{43}.   In this
most traditional approach SUSY is broken in a hidden sector and the scale of SUSY breaking is very
large of order
$\Lambda\sim\sqrt{G^{-1/2}_F M_P}$  where
$M_P$ is the Planck mass. But since the hidden sector only communicates with the visible sector
through gravitational interactions the splitting of the SUSY multiplets is much smaller, in the TeV
energy domain, and the Goldstino is practically decoupled. 
But alternative mechanisms of SUSY breaking are also being considered
\cite{hall}. In one alternative scenario the (not so
much) hidden sector is connected to the visible one by ordinary gauge interactions. As these are much
stronger than the gravitational interactions, $\Lambda$ can be much smaller, as low as 10-100
TeV. It follows that the Goldstino is very light in these models (with mass of order or below 1 eV
typically) and is the lightest, stable SUSY particle, but its couplings are observably large. The radiative
decay of the lightest neutralino into the Goldstino leads to detectable photons. The signature of photons comes
out naturally in this SUSY breaking pattern: with respect to the MSSM, in the gauge mediated model there are typically
more photons and less missing energy. The main appeal of gauge mediated models is a better protection against
flavour changing neutral currents. In the gravitational version even if we accept that gravity leads to
degenerate scalar masses at a scale near $M_{P}$ the running of the masses down to the weak scale can
generate mixing induced by the large masses of the third generation fermions \cite{hall}. More recently it has been
pointed out that there are pure gravity contributions to soft masses that arise from gravity theory
anomalies \cite{hall}. In the assumption that these terms are dominant the associated spectrum and phenomenology has been
studied. In this case gaugino masses are proportional to gauge coupling beta functions, so that the gluino is much heavier
than the electroweak gauginos, and the wino is most often the lightest SUSY particle. 

The MSSM \cite{43} is a completely specified,
consistent and computable theory. There are too many parameters to attempt a direct fit of the electroweak precision data to
the most general framework. But we can consider two significant limiting cases: the "heavy" and the
"light" MSSM.

	The "heavy" limit corresponds to all s-particles being sufficiently massive, still within the limits
of a natural explanation of the weak scale of mass. In this limit a very important result holds: for what concerns the
precision electroweak tests, the MSSM predictions tend to reproduce the results of the SM with a light Higgs, say $m_H\sim$
100 GeV. So if the masses of SUSY partners are pushed at sufficiently large values the same quality of fit as for the SM is
guaranteed. 

	In the "light" MSSM option some of the superpartners have a relatively small mass, close to their
experimental lower bounds. In this case the pattern of radiative corrections may sizeably deviate from
that of the SM.. The potentially largest effects occur in vacuum polarization amplitudes and/or the
$Z\rightarrow b\bar b$ vertex. Since no sign of deviations from the SM is seen in the data and no light SUSY partners
have been found at LEP2 or at the Tevatron, the "light" case can no more be that light.

According to the prevailing view at present, the large scale structure of particle physics consists of a unified theory
at
$M\approx M_{GUT} \sim M_P$ and a low energy effective theory valid at and above the weak scale of energy.  The lagrangian
density  of the low energy  effective theory , after integrating out all very heavy degrees of
freedom,  consists of a set of operators of dimension non larger than 4, that correspond to the renormalisable part, plus a
set of higher dimension, non renormalisable, operators. Schematically, we have:
\beq
{\cal L}=\mu^2 \phi^2+m \bar\psi \psi + g \bar\psi iD\llap{$/$} \psi+\lambda \phi^4+......+
\frac{\lambda_5}{M}\bar\psi \psi \phi \phi+\frac{\lambda_6}{M^2}\bar\psi \psi \bar\psi \psi+....\label{eff}
\eeq
Indicatively, we have shown a number of typical terms of dimension 2 (boson masses), 3 (fermion masses), 4
(renormalisable interactions) plus examples of operators of higher dimension, 5 and 6. Due to the very
large scale of energy where the really fundamental theory applies, the conditions on the low energy effective theory
are severe. First, the dimension $\leq4$ part must be renormalisable. This is a minimum requirement in order to have a closed,
consistent and predictive description of the dynamics after the presence of the very high cut off has been hidden inside
renormalised masses and couplings. But this is not enough because the dependence of masses and couplings from the cut off must
be reasonable in order to avoid the necessity of immense fine tuning. For this to be true additional conditions must be
satisfied. The coupling in front of each
operator, in absence of specific reasons, should be proportional to the large cut off $M$ raised to the power d fixed by
dimensions. For example, $\mu^2$ should be proportional to $M^2$. In the SM there is no symmetry reason why this should
not be the case. So boson masses, like the W and Z masses, should be of order M. This the hierarchy problem. In
supersymmetric extensions of the SM $\mu^2$ is instead of order the mass splittings of SUSY multiplets, because in the
limit of exact SUSY symmetry there are no quadratic divergences (in presence of boson-fermion symmetry the stronger
bosonic divergences must disappear, in order that bosonic and fermionic divergences can both be logaritmic). For fermions
$m$ is not of order $M$ but of order
$v\log{M}$ because the divergences in the fermionic sector are always at most logaritmic. Also, chiral symmetry ensures
that if you start from zero masses the quantum corrections to $m$ must vanish. Once supersymmetry
or some other stabilising mechanism is introduced, the renormalisable part of the lagrangian is sufficiently insensitive
to the presence of the very large cut off $M$. The additional non renormalisable terms are suppressed by powers of $M$.
At energies of order $v$, the electro-weak scale, their effects are proportional to $(v/M)^d$, $d=1,2,...$, hence very
small.

\subsection{Hints from Experiment}
\subsubsection{Unification of Couplings}

At present the most direct
phenomenological evidence in favour of supersymmetry is obtained from the unification of couplings in GUTs.
Precise LEP data on $\alpha_s(m_Z)$ and $\sin^2{\theta_W}$ confirm what was already known with less accuracy:
standard one-scale GUTs fail in predicting $\sin^2{\theta_W}$ given
$\alpha_s(m_Z)$ (and $\alpha(m_Z)$) while SUSY GUTs \cite{ross} are in agreement with the present, very precise,
experimental results. According to a recent analysis, if one starts from the known values of
$\sin^2{\theta_W}$ and $\alpha(m_Z)$, one finds for $\alpha_s(m_Z)$ the results:
\bea
		\alpha_s(m_Z) = 0.073\pm 0.002 ~~~~~      	(\rm{Standard~ GUTs})\nonumber \\	
		\alpha_s(m_Z) = 0.129\pm0.010~~~~~  (\rm{SUSY~ GUTs})
\label{onehundredthirty}
\eea
to be compared with the world average experimental value $\alpha_s(m_Z)$ =0.119(3).

\subsubsection{Dark Matter}

There is solid astrophysical and cosmological evidence \cite{prim}, that most of the matter in the universe
does not emit electromagnetic radiation, hence is "dark". Some of the dark matter must be baryonic but most of it must
be non baryonic. Non baryonic dark matter can be cold or hot. Cold means non relativistic at freeze out, while hot is
relativistic. There is general consensus that most of the non baryonic dark matter must be cold dark matter. A
couple of years ago the most likely composition was quoted to be around $80\%$ cold and $20\%$ hot. At present it appears
that the need of a sizeable hot dark matter component is more uncertain. In the last few years great progress has been made
in the experimental determination of fundamental cosmological parameters. The Hubble constant has been measured, also
using the Hubble telescope, ($H_0 = 65\pm8 km~s{-1}~Mpc^{-1}$). There is growing experimental evidence (for example, from the
supernovae distribtion vs redshift) of the presence of a cosmological constant component in
$\Omega=\Omega_m+\Omega_{\Lambda}$. Here
$\Omega$ is the total matter-energy density in units of the critical density, $\Omega_m $ is the matter component
(dominated by  non baryonic cold dark matter) and $\Omega_{\Lambda}$ is the cosmological component. $\Omega_m$ is
extimated reliably, for example from the mass distribution at large distances, measured by gravitational lensing, which gives
$\Omega_m \approx 0.35$. Inflationary theories strongly favour
$\Omega=1$ which is consistent with present data (in particular the beautiful new data on the position of the first acoustic
peak by Boomerang and Maxima). At present, still within large uncertainties, the approximate composition is indicated to
be
$\Omega_m\sim 0.35$ and
$\Omega_{\Lambda}\sim 0.65$ (baryonic dark matter from big bang nucleosynthesis gives $\Omega_b\sim 0.05$). 

The implications for particle physics is that certainly there must exist a source of cold dark matter. By far the
most appealing candidate is the neutralino, the lowest supersymmetric particle, in general a superposition of
photino, Z-ino and higgsinos. This is stable in supersymmetric models with R parity conservation, which are the
most standard variety for this class of models (including the MSSM). A
neutralino with mass of order 100 GeV would fit perfectly as a cold dark matter candidate. Another common
candidate for cold dark matter is the axion, the elusive particle associated to a possible solution of the strong
CP problem along the line of a spontaneously broken Peccei-Quinn symmetry. To my knowledge and taste this
option is less plausible than the neutralino. One favours supersymmetry for very diverse conceptual and
phenomenological reasons, as described in the previous sections, so that neutralinos are sort of standard by now.
For hot dark matter, the self imposing candidates would be neutrinos. If we demand a density fraction
$\Omega_{\nu}\sim 0.1$ from neutrinos, the maximum which is allowed by observations,then it turns out that the sum of stable
neutrino masses should be around 5 eV.

\subsubsection{Neutrino Masses}

Recent data \cite{os} from Superkamiokande have provided a more solid experimental basis for neutrino
oscillations as an explanation of the atmospheric neutrino anomaly. In addition the solar neutrino deficit,
observed by several experiments, is also probably an indication of a different sort of neutrino oscillations. Results
from the laboratory experiment by the LSND collaboration, not confirmed by KARMEN, can also be considered as a
possible indication of yet another type of neutrino oscillation.  Neutrino oscillations imply neutrino masses. The extreme
smallness of neutrino masses in comparison with quark and charged lepton masses indicate a different nature of neutrino
masses, linked to lepton number violation and the Majorana nature of neutrinos. Thus neutrino masses provide a window on the
very large energy scale where lepton number is violated and on GUTs. The new experimental evidence on
neutrino masses could also give an important feedback on the problem of quark and charged lepton masses, as all these
masses are possibly related in GUTs. In particular the observation of a nearly maximal mixing angle for
$\nu_{\mu}\rightarrow \nu_{\tau}$ is particularly interesting. Perhaps also solar neutrinos may occur with
large mixing angle. At present solar neutrino mixings can be either large or very small, depending on which particular
solution will eventually be established by the data. Large mixings are very interesting because a first guess was in
favour of small mixings in the neutrino sector in analogy to what is observed for quarks. If confirmed, single or double
maximal mixings can provide an important hint on the mechanisms that generate neutrino masses.

From a strict minimal standard model point
of view neutrino masses could vanish if no right handed neutrinos existed (no Dirac mass) and lepton number was
conserved (no Majorana mass). In GUTs both these assumptions are violated. The right handed neutrino is required in all
unifying groups larger than SU(5). In SO(10) the 16 fermion fields in each family, including the right handed neutrino,
exactly fit into the 16 dimensional representation of this group. This is really telling us that there is something in
SO(10)! The SU(5) alternative in terms of $\bar 5+10$, without a right handed neutrino, is certainly less elegant. The
breaking of
$|B-L|$, B and L is also a generic feature of GUTs. In fact, the see-saw mechanism \cite{af} explains
the smallness of neutrino masses in terms of the large mass scale where $|B-L|$ and L are violated. Thus, neutrino
masses, as would be proton decay, are important as a probe into the physics at the GUT scale.

Oscillations only determine squared mass differences and not masses. If in addition to solar and atmospheric neutrino
oscillations  also the LSND evidence will be confirmed, then one would need to add a fourth sterile neutrino (i.e. without
weak interactions, to avoid the LEP veto against additional light weakly interacting neutrinos). This is because oscillation
frequencies determine squared mass differences and with three masses there are only two independent differences. However,
sterile neutrinos are at present disfavoured both from atmospheric and solar neutrino oscillation observations. Thus in the
following we assume that the LSND evidence for a third oscillation frequency will disappear and we restrict to three
neutrinos. In terms of our labelling of  masses the two frequencies are given by $\Delta_{sun}\propto m^2_2-m^2_1$ and
$\Delta_{atm}\propto m^2_3-m^2_{1,2}$.

Neutrino oscillations only determine differences of squared masses and not the absolute mass scale. the case of three almost
perfectly degenerate neutrinos is the only one that could in principle accomodate neutrinos as hot dark matter together with
solar and atmospheric neutrino oscillations. According to our previous discussion, the common mass should be around 1-2 eV.
The solar frequency ($\Delta m^2_{sun}\sim 10^{-5}-10^{-10}~eV^2$, depending on which solution is finally established) could be
given by a small 1-2 splitting, while the atmospheric frequency could be given by a still small but much larger 1,2-3 splitting
($\Delta m^2_{atm}\sim 3~10^{-3}~eV^2$). A strong constraint arises in the degenerate case from neutrinoless double beta decay
which requires that the ee entry of
$m_{\nu}$ must obey
$|(m_{\nu})_{11}|\leq 0.2-0.5~{\rm eV}$ \cite{af}. It has been observed that this bound can only be 
satisfied if
double maximal mixing is realized, i.e. if also solar neutrino oscillations occur with nearly maximal mixing.
We have mentioned that it is not at all clear at the moment that a hot dark matter component is really
needed \cite{prim}. However the only reason to consider the fully degenerate solution is 
that it is compatible
with hot dark matter.
Note that for degenerate masses with $m\sim 1-2~{\rm eV}$ we need a relative splitting $\Delta m/m\sim
\Delta m^2_{atm}/2m^2\sim 10^{-3}$ and a much smaller one for solar neutrinos. It is not simple
to imagine a natural mechanism compatible with unification and the see-saw mechanism to arrange such a
precise near symmetry.

If neutrino masses are smaller than for cosmological relevance, we can have the hierarchies $|m_3| >> |m_{2,1}|$
or $|m_1|\sim |m_2| >> |m_3|$. We prefer the first case, because for quarks and leptons one
mass eigenvalue, the third generation one, is largely dominant. Thus the dominance of $m_3$ for neutrinos
corresponds to what we observe for the other fermions.  In this case, $m_3$ is determined by the atmospheric
neutrino oscillation frequency to be around $m_3\sim0.05~eV$. By the see-saw mechanism $m_3$ is related to some
large mass M, by $m_3\sim m^2/M$. If we identify m with either the Higgs vacuum expectation value or the top mass
(which are of the same order), as suggested for third generation neutrinos by GUTs in simple SO(10)
models, then M turns out to be around $M\sim 10^{15}~GeV$, which is consistent with the connection with GUTs.

A lot of attention \cite{af} is being devoted to the
problem of a natural explanation of the observed nearly maximal mixing angle for atmospheric
neutrino oscillations and possibly also for solar neutrino oscillations, if explained by vacuum
oscillations. Large mixing angles are somewhat unexpected because
the observed quark mixings are small and the quark, charged lepton and neutrino mass matrices are to
some extent related in GUT's. There must be some special interplay between the neutrino Dirac
and Majorana matrices in the see-saw mechanism in order to generate maximal
mixing. It is hoped that looking for a natural explanation of large neutrino mixings can lead us to decripting
some interesting message on the physics at the GUT scale.

\subsubsection{Baryogenesis}

Baryogenesis is interesting because it could occur at the weak
scale \cite{rev} but not in the SM. For baryogenesis one needs the three famous Sakharov conditions: B
violation, CP violation and no termal equilibrium. In principle these conditions could be verified in the SM. B is
violated by instantons when kT is of the order of the weak scale (but B-L is conserved). CP is violated by the CKM
phase and sufficiently marked out of equilibrium conditions could be realised during the electroweak phase transition. So
the conditions for baryogenesis  at the weak scale in the SM appear superficially to be present.
However, a more quantitative analysis \cite{rev}, shows that baryogenesis is not possible
in the SM because there is not enough CP violation and the phase transition is not sufficiently strong first order,
unless
$m_H<80~GeV$, which is by now completely excluded by LEP. However, it is interesting that baryogenesis at the weak scale is not
yet excluded in SUSY extensions of the SM. In particular, in the MSSM there are additional sources of CP violations and the
bound on $m_H$ is modified by a sufficient amount by the presence of scalars with large couplings to the Higgs sector,
typically the s-top. What is required is that
$m_h\sim 80-110~GeV$, a s-top not heavier than the top quark and, preferentially, a small
$\tan{\beta}$. This possibility has become more and more marginal with the progress of the LEP2 running.

If baryogenesis at the weak scale is excluded by the data it can occur at or just below the
GUT scale, after inflation. But only that part with
$|B-L|>0$ would survive and not be erased at the weak scale by instanton effects. Thus baryogenesis at $kT\sim
10^{12}-10^{15}~GeV$ needs B-L violation at some stage like for $m_\nu$, if neutrinos are Majorana particles. The two
effects could be related if baryogenesis arises from leptogenesis then converted into baryogenesis by
instantons \cite{buch}. Recent results on neutrino masses are compatible with this elegant possibility. Thus the case of
baryogenesis through leptogenesis has been boosted by the recent results on neutrinos.

\section{Status of the Search for the Higgs and for New Physics}

The LEP2 programme has started in the second part of 1995. The the total center of mass energy was gradually increased up to
$208~GeV$. The main goals of LEP2 are the search for the Higgs and for new particles, the measurement of
$m_W$ and the investigation of the triple gauge vertices
$WWZ$ and $WW\gamma$
\cite{lep2}. 

Concerning the Higgs, the present limits (summer '00) obtained by the LEP collaborations , are, for the SM Higgs, $m_H\gappeq
113 GeV$ and for the lightest MSSM Higgs, $m_h\gappeq 90 GeV$. To understand the significance of these limits we recall the
theoretical bounds on the Higgs mass. 
	
It is well known that in the SM with only one Higgs doublet a lower limit on
$m_H$ can be derived from the requirement of vacuum stability. This criterium is equivalent to demand that the coupling
$\lambda$ of the quartic term $\lambda (\phi \dagger \phi)^2$ does not become negative while running from the weak
scale up to the scale $\Lambda$. The initial value of $\lambda$ at the weak scale increases with $m_H^2$, while the
derivative, for $m_H$ near the limit, is dominated, for a not too heavy Higgs, by the top quark term which is large and
negative. The value of the limit is a function of
$m_t$ and of the energy scale
$\Lambda$ where the model breaks down and new physics appears.  
If one requires that
$\lambda$ remains positive up to $\Lambda = 10^{15}$--$10^{19}$~GeV, then the resulting bound on $m_H$ in the SM with
only one Higgs doublet is given by:
\begin{equation} m_H > 135 + 2.1 \left[ m_t - 174.3 \right] - 4.5~\frac{\alpha_s(m_Z) - 0.119}{0.006}~.
\label{25h}
\end{equation}
It follows that he discovery of a SM-like Higgs particle at
LEP2, or $m_H\lappeq 115~GeV$, would imply that the SM breaks down at a scale
$\Lambda$ of the order of $\lappeq 100~TeV$. Note, however, that the lower bound is invalidated if more than one single Higgs
doublet exists: for more doublets the limit applies to some average mass and not to the lightest Higgs particle.

Similarly an upper bound on $m_H$ (with mild dependence
on $m_t$) is obtained from the requirement that up to the scale $\Lambda$ no Landau pole appears. The upper limit
on the Higgs mass in the SM is important to guarantee the success of the LHC as an accelerator designed to solve
the Higgs problem.  In
fact, for large Higgs masses, the initial value of $\lambda$ is large and the derivative of $\lambda$ is positive, because
the positive $\lambda$ term (the $\lambda \phi^4$ theory is not asymptotically free!) overwhelms the top Yukawa negative
contribution. As a consequence the coupling $\lambda$ tends to infinity (the Landau pole) at some finite scale.
The upper limit on $m_H$ has been studied not only in perturbation theory but also using lattice simulations of the Higgs
sector in the region near the pole which is non perturbative. For
$m_t\sim 175~GeV$ one finds
$m_H\lappeq 180~GeV$ for $\Lambda\sim M_{GUT}-M_{Pl}$ and $m_H\lappeq 0.5-0.8~TeV$ for $\Lambda\sim
1~TeV$. Thus, in conclusion  \cite{hr}, if the SM holds up
to $\Lambda \sim M_{GUT}$ or $M_{Pl}$, then, 
 for
$m_t \sim$ 174~GeV, only a small range of values for $m_H$ is allowed, $130 < m_H <~\sim 200$~GeV. 

A particularly
important example of theory where the lower bound is violated, is the
MSSM, which we now discuss. As is well known \cite{43}, in the MSSM there are two Higgs doublets, which implies three
neutral physical Higgs particles and a pair of charged Higgses. The lightest neutral Higgs, called $h$, should be
lighter than
$m_Z$ at tree-level approximation. However, radiative corrections increase the $h$ mass by a term
proportional to $m^4_t$ and  logarithmically dependent on the stop mass. Once the radiative corrections are taken into
account the $h$ mass still remains rather small: for $m_t = 174~GeV$ one finds the limit $m_h \lappeq 130~GeV$ 
(valid for all values of $tg\beta$ and saturated at large $tg\beta$). Actually one can well expect that $m_h$ is sizeably
below the bound if  $tg\beta$ is small, $tg\beta = v_{up}/v_{down} < 10$). LEP2 is progressively excluding a part of the small
$\tan{\beta}$ domain. If no Higgs is found at LEP the domain $\tan{\beta}\lappeq 2-8$  will be  excluded, depending on the
value of other MSSM parameters.
 By now most of
the discovery potential of LEP2 for supersymmetric particles has been deployed. For example, the limit on the chargino mass
was
about $45~GeV$ after LEP1 and is now about $m_{\chi^+}\lappeq103~GeV$, apart from exceptional regions of the MSSM parameter
space. The lightest neutralino mass limit is around $m_{\chi^0}\lappeq40~GeV$. The region of the MSSM parameter space that
has been by now excluded by LEP is a very important one. The low $tg\beta$ solution was appealing in many respects. Some more
constrained forms of the model, like the supergravity version, where degenerate scalar masses and gaugino masses are assumed
at the GUT scale, are by now disfavoured. With no discovery of the Higgs and SUSY at LEP the case for the MSSM certainly
becomes less natural, and even less natural become the gauge mediated models.

	An important competitor of CERN is the Tevatron collider. In 2001 the Tevatron will start RunII with the purpose
of collecting a few $fb^{-1}$ of integrated luminosity at $2~TeV$. The competition is especially on the search of
new particles and the Higgs, but also on
$m_W$ and the triple gauge vertices. For example, for supersymmetry,  LEP2 was strong on Higgses, charginos, neutralinos and
sleptons while the Tevatron is superior for gluinos and squarks, . There are plans for RunIII to start in
$\gappeq 2004$ with the purpose
of collecting of the order  $5~fb^{-1}$ of integrated luminosity per year. If so the Tevatron could also hope to find
the Higgs before the LHC if the Higgs mass is close to the LEP2 range.

\section{Conclusion}

Today in particle physics we follow a double approach: from above and from below. From above there are, on the theory
side, quantum gravity (that is superstrings), GUT theories and cosmological scenarios. On the experimental side there
are underground experiments (e.g. searches for neutrino oscillations and proton decay), cosmic ray
observations, space experiments (like COBE, Boomerang, Maxima etc), cosmological observations and so on. From below, the main
objectives of theory and experiment are the search of the Higgs and of signals of particles beyond the Standard Model
(typically supersymmetric particles). Another important direction of research is aimed at the exploration of the flavour
problem: study of CP violation and rare decays. The general expectation is that new physics is close by and that should be
found very soon if not for the complexity of the necessary experimental technology that makes the involved time scale painfully
long.

\end{document}